\documentclass[useAMS,usenatbib]{mn2e}
\usepackage{graphicx}
\usepackage{cleveref}
\usepackage{float}
\usepackage{amsmath}
\usepackage{subfigure}
\usepackage{wrapfig}
\usepackage{rotating}
\usepackage{amssymb}

\title[Shape asymmetry]{Shape asymmetry: a morphological indicator for automatic detection of galaxies in the post-coalescence merger stages.}

\author[M. M.~Pawlik et al. ]{M. M.~Pawlik$^{1}$, V.~Wild$^{1,2}$,  C. J.~Walcher$^{3}$, P. H.~Johansson$^{4}$, C.~Villforth$^{1}$,
\newauthor K.~Rowlands$^{1}$, J.~Mendez-Abreu$^{1}$ and T.~Hewlett$^{1}$ \\ 
$^{1}$School of Physics and Astronomy, University of St Andrews, North Haugh, St Andrews, KY16 9SS, U.K. (SUPA)\\
$^{2}$Institute for Astronomy, University of Edinburgh, Royal Observatory, Blackford Hill, Edinburgh, EH9 3HJ, U.K. (SUPA)\\
$^{3}$Leibniz-Institut f\"ur Astrophysik Potsdam (AIP), An der Sternwarte 16, D-14482 Potsdam, Germany\\
$^{4}$Department of Physics, University of Helsinki, Gustaf H\"allstr\"omin katu 2a, FI-00014 Helsinki, Finland}

\makeatletter

\newcommand{\Rmnum}[1]{\expandafter\@slowromancap\romannumeral #1@}
\makeatother

\begin{document}

\parindent=1cm

\date{}

\pagerange{\pageref{firstpage}--\pageref{lastpage}} \pubyear{0000}

\maketitle

\label{firstpage}

\begin{abstract}

We present a new morphological indicator designed for automated recognition of galaxies with faint asymmetric tidal features suggestive of an ongoing or past merger. We use the new indicator, together with preexisting diagnostics of galaxy structure to study the role of galaxy mergers in inducing (post-)starburst spectral signatures in local galaxies, and investigate whether (post-)starburst galaxies play a role in the the build up of the `red sequence'. Our morphological and structural analysis of an evolutionary sample of 335 (post-)starburst galaxies in the SDSS DR7 with starburst ages $0<t_{SB}<0.6$ Gyr, shows that $45\%$ of galaxies with young starbursts ($t_{SB}<0.1$ Gyr) show signatures of an ongoing or past merger. This fraction declines with starburst age, and we find a good agreement between automated and visual classifications. 
The majority of the oldest (post-)starburst galaxies in our sample ($t_{SB}\sim0.6$ Gyr) have structural properties characteristic of early-type disks and are not as highly concentrated as the fully quenched galaxies commonly found on the `red sequence' in the present day Universe. This suggests that, if (post-)starburst galaxies are a transition phase between active star-formation and quiescence, they do not attain the structure of presently quenched galaxies within the first 0.6 Gyr after the starburst. \end{abstract}


\section{Introduction}

The contrast between `blue-cloud' and `red-sequence' galaxies has long been known and examined in many studies both in the local Universe \citep[e.g.][]{Strateva et al. 2001, Kauffmann et al. 2003b, Baldry et al. 2006} and at higher redshifts \citep[e.g.][]{Bell et al. 2004, Bundy et al. 2005}. The star-forming blue-cloud galaxies tend to have late-type morphologies, while the quiescent systems populating the red sequence are predominantly early-type. The origin of this bimodality remains unclear and has become one of the major conundrums in the field of extragalactic astronomy. As the stellar mass contained within the red sequence has doubled since $z\sim1$, contrary to that of the blue cloud, it is thought that blue galaxies can migrate to the red sequence once their star formation has quenched \citep{Bell et al. 2004, Faber et al. 2007, Arnouts et al. 2007}. While several mechanisms responsible for the star-formation quenching have been studied \citep{Gunn Gott 1972, Balogh Morris 2000, Birnboim Dekel 2003, Di Matteo Springel Hernquist 2005, Cox et al. 2006, Hopkins et al. 2007, Martig et al. 2009, Johansson Naab Ostriker 2012}, the emerging consensus is that the transformation between star-forming and quiescent galaxies can occur through either a fast or a slow channel. The slow mode involves the secular evolution of galaxies, without the presence of external triggering mechanisms, where star-formation fades gradually as the gas supply is used up \citep{Noeske et al. 2007, Cortese Hughes 2009, Schawinski et al. 2014}. This work's focus is on the fast route, supported by the scarcity of galaxies in the `green valley', stretching between the blue cloud and the red sequence in the colour-magnitude diagrams. Fast quenching is commonly associated with an escalated gas consumption during a merger-induced starburst, followed by removal of the remaining gas supply by subsequent stellar and/or AGN feedback \citep{Sanders et al. 1988, Hopkins et al. 2006, Faber et al. 2007, Johansson Naab Burkert 2009, Yesuf et al. 2014}. However, a recent study of local galaxies with signatures of a past starburst by \citet{Rowlands et al. 2015} has shown that they preserve substantial gas reservoirs which can potentially fuel star formation for an extended period of time, making the `fast' nature of this quenching scenario debatable. Further study of the possible quenching mechanisms is required, and galaxies that show signatures of a recent starburst (post-starburst galaxies) as well as those that have undergone a recent merger (post-merger galaxies) may provide a unique insight to this `fast' transition phase between the star-forming and the quiescent mode of galaxy evolution \citep{Tran et al. 2004, Kaviraj et al. 2007}. 

Post-starburst (PSB) galaxies (also known as `k+a' galaxies) have spectra featuring unusually strong Balmer lines in absorption - an attribute of A stars. Such features are thought to be a signature of a recent ($\sim~1$ Gyr) truncation following either a short burst \citep{Dressler Gunn 1983, Nolan Raychaudhury Kaban 2007} or normal-mode star formation activity within a galaxy \citep{Poggianti et al. 1999}. Distinguishing between the two events requires careful modelling, which shows that very strong Balmer lines are inconsistent with anything other than a recent starburst \citep{Balogh et al. 2005, Wild et al. 2007, von der Linden et al. 2010}.
First discovered over 30 years ago \citep{Dressler Gunn 1983}, PSB galaxies are found in various regions of the Universe: locally, they are rare and populate mainly the field and poor groups \citep{Zabludoff et al. 1996, Blake et al. 2004} but their incidence increases at higher redshifts, where they seem to lie predominantly in clusters (e.g. \citealt{Dressler et al. 1999, Ma et al. 2008, Poggianti et al. 2009} but see \citealt{Balogh et al. 1999} for a contradictory conclusion). Whether PSB galaxies are a link between the star-forming and quiescent phases of galaxy evolution is still an ongoing debate. Results of some studies support this claim: at $0.5<z<1$, the mass flux through the PSB phase is comparable to the mass growth of the red sequence \citep{Wild et al. 2009}, and  quenched PSB galaxies tend to be early-type \citep{Wong et al. 2012, Mendel et al. 2013}. However, other studies suggest the opposite: the incidence of PSB galaxies at $0.5<z<1$ is inconsistent with a major channel for the red sequence growth in clusters \citep{de Lucia et al. 2009, Dressler et al. 2013}, and the remaining gas reservoirs up to $0.5-1$ billion years after the starburst suggest a non-quenched state for local PSB galaxies \citep{Zwaan et al. 2013, Rowlands et al. 2015}. The contradictory conclusions of the different studies may partially come from their differing selection techniques used to identify PSB signatures in galaxies, as well as the different environments and redshifts that each study focuses on. Clearly further study of this interesting class of galaxies is warranted. 

The origin of PSB signatures in galaxies has not yet been constrained. A possible mechanism, favoured by numerical simulations (e.g. \citealt{Barnes Hernquist 1991, Barnes Hernquist 1996, Bekki et al. 2001, Bekki et al. 2005, Hopkins et al. 2006, Hopkins et al. 2008, Snyder et al. 2011}), is gas-rich major mergers, which can induce starbursts strong enough to rapidly consume a significant amount of the galaxy's gas reservoir \citep{Di Matteo Springel Hernquist 2005}, and can transform late-type galaxies into early types both structurally and kinematically \citep{Toomre Toomre 1972, Naab Burkert 2003}. Mergers of galaxies have been observed for over 50 years \citep{Vorontsov-Velyaminov 1959, Arp 1966, Schweizer 1982, Schweizer et al. 1983, Ellison et al. 2013} and their connection with PSB spectral signatures in galaxies has been investigated in some previous studies but with no definite conclusion so far. The fact that many local PSB galaxies lie in the field and poor groups suggests that they could be merger remnants \citep{Zabludoff et al. 1996, Quintero et al. 2003, Balogh et al. 2005}, and the signs of morphological disturbance found in many PSB galaxies provides strong evidence supporting this claim (\citealt{Zabludoff et al. 1996, Blake et al. 2004, Goto 2005, Yang et al. 2008}). Nonetheless, results reported by some other studies suggest otherwise, for example: not all PSB galaxies show signs of an interaction \citep{Zabludoff et al. 1996, Blake et al. 2004, Yang et al. 2008} and the redshift-evolution of the PSB number density found by \citet{Wild et al. 2009} is $\sim100$ times that of the major-merger rate estimated by \citet{de Ravel et al. 2009}. Furthermore, \citet{Dressler et al. 2013} argue that the relative incidence of starburst, post-starburst, passive and star-forming galaxies found in five rich clusters suggests that less disruptive events, like minor mergers and accretion, are at play. 

The role of galaxy mergers in quenching of star-formation in galaxies is not well understood and it remains challenging to constrain due to the large variability of merger signatures and the resulting difficulties in defining robust sample selection criteria. Common methods include selection of close pairs \citep[i.e. pre-merger stages, ][]{Zepf Koo 1989, Patton et al. 2002} or probing visual merger signatures in galactic structure. The latter approach can be done by means of either visual inspection or quantitative approach involving, for example: rotational asymmetry \citep[e.g.][]{Rix Zaritsky 1995, Shade et al. 1995, Abraham et al. 1996, Conselice et al. 2000, Villforth et al. 2014}, lopsidedness \citep{Reichard et al. 2008, Reichard et al. 2009} the Gini$-M_{20}$ parameter space \citep[e.g.][]{Lotz et al. 2004} or the recently introduced median-filtering of high-resolution images for detection of double nuclei \citep{Lackner et al. 2014}. Such measures are useful for identifying galaxies in early stages of a merger, particularly prior to the coalescence of the galaxy nuclei but, they are less suitable for tracing low surface-brightness post-merger signatures.
While visual classification offers great reliability and have recently been extended to large galaxy samples (The Galaxy Zoo Project, \citealt{Lintott et al. 2008, Lintott et al. 2011}; see \citealt{Darg et al. 2010} for merger morphology), the need for a more efficient, automated selection of galaxies in post-coalescence stages of a merger increases in today's era of extensive surveys. An automated approach also provides a more quantitative and easily reproducable description of galaxy morphology than visual classification.

In this paper we introduce a new morphological indicator designed to probe the asymmetric structures in the outskirts of galaxies in the late stages of a merger. Through comparison with visual classification of galaxy images, we show that the indicator performs remarkably well in detecting faint signatures of tidal disruption in starburst and post-starburst galaxies. We use the new measure to identify post-merger (or, in few cases, ongoing-merger) candidates in a sample of 335 local galaxies with bulges, with strongest starburst and post-starburst signatures in the Sloan Digital Sky Survey (SDSS, \citealt{Abazajian et al. 2009}), selected using a sophisticated technique introduced by \citet{Wild et al. 2007}, and to investigate the evolution of the morphological disturbance in these galaxies as a function of the starburst age. We also use several standard measures found in literature to characterise the structure of galaxies evolving through the starburst and post-starburst phase and compare them with those characteristic of galaxies residing on the present-day red sequence.

This paper is structured as follows: Section 2 contains a description of samples and their selection criteria; Section 3 - the methodology; Section 4 - calibration and testing; Section 5 - the results; Section 6 - a discussion; Section 7 - the summary of conclusions. We adopt a cosmology with $\Omega_{m}$ = 0.30, $\Omega_{\Lambda}$ = 0.70 and H$_{0}$ =70kms$^{-1}$Mpc$^{-1}$.


\section{Sample selection}\label{sec:samples}

The following samples of galaxies were selected from the SDSS DR7 spectroscopic catalogue.

\begin{figure}
  \centering  
  \includegraphics[width=\columnwidth]{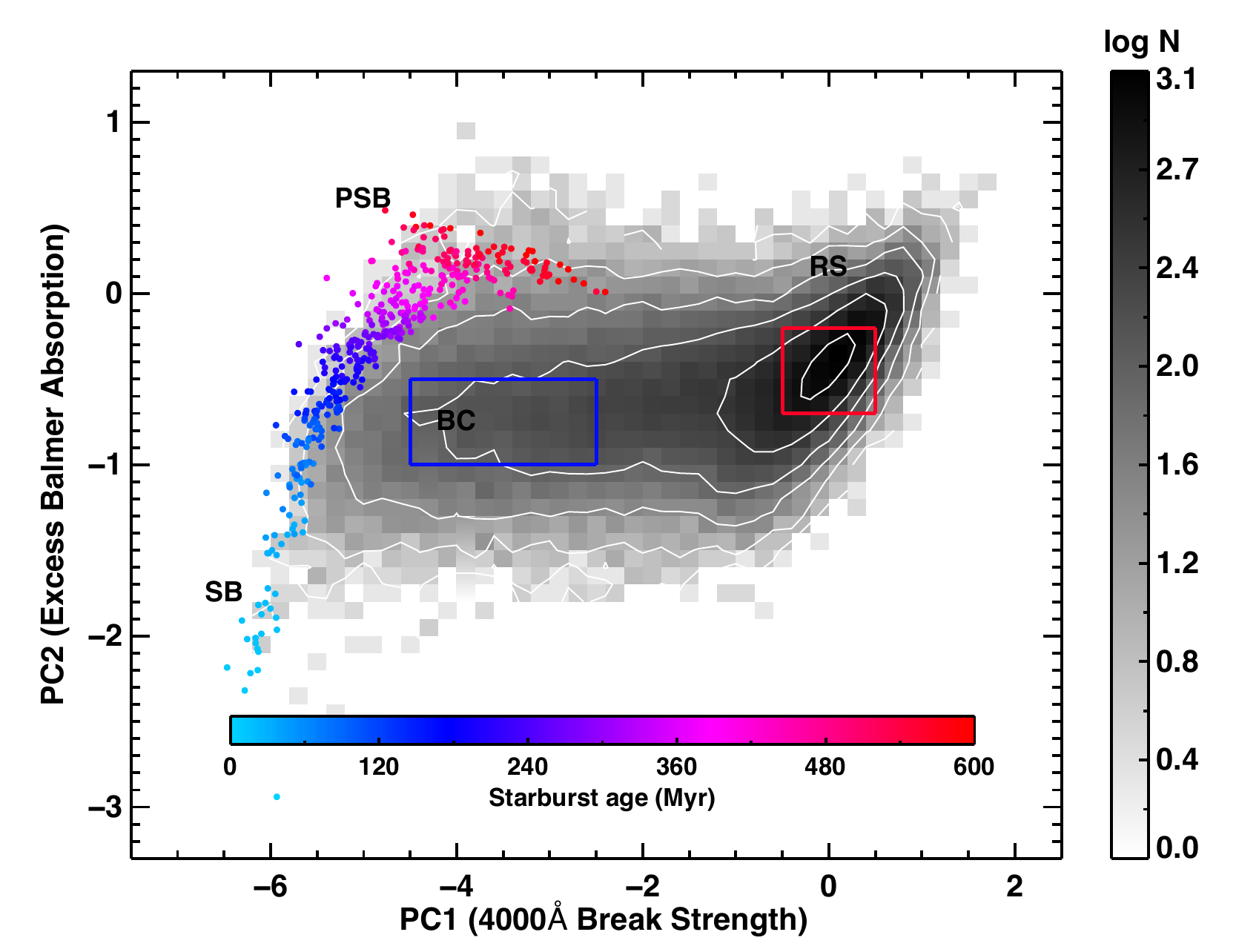}
  \caption{Distribution of the 70,000 local bulge-galaxies selected from SDSS DR7 in the PC1-PC2 spectral index space (greyscale). Galaxies with the strongest (post-)starburst signatures in the sample are overplotted in colour. Different regions of the spectral index space contain galaxies with different recent tar formation histories. These are indicated by the acronyms: SB, PSB, RS and BC, and correspond to: starburst, post-starburst, red sequence and blue cloud, respectively. The blue and red boxes mark the region used for selection of the control sample of blue-cloud and red-sequence galaxies, respectively.}
    \label{fig:vivplot}
\end{figure}

\subsection{The test sample}\label{sec:test_sample}

The test sample consists of 70 local galaxies at $0.01<z<0.07$ with SDSS Petrosian magnitude within the range: $14.5<m_{r}<17.7$. The sample was selected to represent systems with various degrees of morphological disturbance, which was determined based on visual inspection of the galaxy images. 
To find galaxies with disrupted morphologies, we pre-selected galaxies with spectroscopic starburst and post-starburst signatures, as further described in Section \ref{sec:psb_sample}. We then visually examined their images in $r$-band as well as the SDSS false colour images for the presence of morphological disturbance and tidal features, pointing to a past (and in a few cases, still ongoing) merger event\footnote{Troughout the text, we refer to both ongoing and past merger events collectively as past merger events or `post-mergers'.}. The final sample of test galaxies (Figures \ref{fig:morphsample1} and \ref{fig:morphsample2}) contains: galaxies with highly disrupted morphologies, tidal features indicating a past major merger (images 1-20); galaxies with moderate morphological disturbance, with no prominent tidal features but slight deviations from regular appearance (images 21-30); galaxies with no signs of morphological disturbance (images 31-40). The notable fraction of ongoing mergers in the sample is a consequence of the selection criteria chosen for purposes of the method testing. The test sample is not representative of the entire population of local PSB galaxies.
Finally, we include in the test sample a control subset of 30 normal galaxies (selected using the above constraints on redshift and magnitude) classified as early- (images 41-50, 61-65) and late-types (images 51-60, 66-70) based on the SDSS \emph{fracDev} parameter with threshold values of 0.99 and 0.1, respectively. The \emph{fracDev} parameter describes the fraction of the total galaxy light fit by the de Vaucouleurs profile (with the total light being represented by the model magnitude, computed from a linear combination of de Vaucouleurs and exponential fits to galaxy light profiles). We split the normal galaxies based on their inclination, into face-on ($b/a>0.5$, images 41-60) and edge-on ($b/a<0.5$, images 61-70) subsets. 

\subsection{The (post-)starburst sample}\label{sec:psb_sample}

In this work we are interested in the whole sequence from starburst to quiescence, and wish to observe the decline in post-merger features that might be expected following the starburst event. We therefore select an evolutionary sample of galaxies at different stages of the starburst/post-starburst phase, which we collectively refer to as the `(post-)starburst' or `SB/PSB' sample. The sample consists of 400 local galaxies which have undergone a strong recent starburst selected from a parent sample of 70,000 galaxies with $0.01<z<0.07$ and spectral SNR/pixel greater than 8 in the $g$-band, from the spectroscopic SDSS DR7 catalog. The galaxies within the parent sample were selected to be bulge dominated, with stellar surface mass density $\mu>$ 3 $\times$10$^{8}$M$_{\sun}$ kpc$^{-2}$ \footnote{$\mu = M_{*}/(2\pi r^{2})$ with $r$ being the physical size in kpc of the radius which contains $50\%$ of the SDSS' $z$-band Petrosian flux and the stellar mass measured from the 5-band SDSS photometry (J. Brinchmann, http://www.mpa-garching.mpg.de/SDSS)}. This is similar to imposing a stellar mass limit of $10^{10}$M$_{\sun}$ \citep{Kauffmann et al. 2003b}. Our reason for using this sample is that it has already been studied in \citet{Wild et al. 2010} to measure a $\sim200$ Myr offset between starburst and accretion onto the central supermassive black hole, and in \citet{Rowlands et al. 2015} to investigate the evolution of their dust and gas contents. The disadvantage of this pre-selection on stellar mass surface density is that galaxies with bulges may undergo stronger starbursts due to gravitational effects, and therefore not be representative of the full merging galaxy population. We intend on extending our analysis to further samples in later work. 

Figure \ref{fig:vivplot} shows the distribution of two spectral indices that describe the shape of the spectrum (4000\AA\ break and Balmer break strength) and Balmer absorption line strength for all 70,000 bulge-dominated galaxies. These spectral features constrain the recent star formation history of the galaxy \citep{Kauffmann et al. 2003a}. The indices are computed within the spectral region $3175$\AA-$4150$\AA\ using a principal component analysis (PCA), which identifies and groups together features that vary simultaneously as the balance of stars in a spectrum changes \citep{Wild et al. 2007}. By combining multiple Balmer absorption lines together with information from the shape of the stellar continuum the PCA provides a much higher signal-to-noise ratio spectral index to identify an excess of A stars compared to the traditionally used H$\beta$ or H$\delta$ absorption lines. In Figure \ref{fig:vivplot}, a large majority of the galaxies are found on the right side of the plot with strong 4000\AA\ break (PC1), characteristic of the quiescent red-sequence galaxies. Lower values of PC1 point to ongoing star formation, therefore the less numerous blue-cloud galaxies appear on the left side of the plot. Starburst galaxies undergo brief episodes of enhanced star formation which reflect in extremely low values of PC1 as well as PC2 (deficit of Balmer absorption and weak 4000\AA\ breaks), due to their stellar populations being dominated by short-lived O/B stars. As the dominant stellar populations within these galaxies change with time elapsed from the starburst event, the galaxy moves to the post-starburst phase, and the Balmer absorption features will increase in strength relative to the 4000\AA\ break \citep{Dressler Gunn 1983, Couch Sharples 1987}. The peak at highest PC2 values is where we find old post-starburst galaxies with dominant populations of A stars and starburst age ($t_{SB}$) $\sim$ 1Gyr. 

To select an evolutionary sample of (post-)starburst galaxies, we selected galaxies from the extreme left of the PC1/2 distribution. These have undergone the strongest recent starburst in the sample, and therefore exhibit the most extreme spectral features. The starburst ages of those galaxies were estimated using Bayesian fitting of \citet{Bruzual Charlot 2003} stellar population synthesis models to the spectral indices. First, starburst ages for  all galaxies with PC1$<-4$ or PC2$>-0.5$ were estimated from the median of the posterior distribution, assuming a star formation history comprising an old stellar population and recent starburst. To identify a statistically complete sample of the strongest (post-)starburst galaxies we selected galaxies with the lowest PC1 value, such that we have 20 galaxies per 30\,Myr time interval, up to a starburst age of 600\,Myr. The reason for this age restriction is that, at older ages, there is a degeneracy between starburst age and burst strength using PC1 and PC2 alone (see \citealt{Wild et al. 2007}). The final sample of 400 SB/PSB galaxies are plotted in Figure \ref{fig:vivplot}, with starburst age coded by colour.

The key difference between this selection method and those used in previous studies of post-starburst galaxies is that we select purely on stellar continuum features and do not remove galaxies with identifiable nebular emission lines. Traditionally galaxies are removed from the post-starburst samples if either H$\alpha$ or [O$\mbox{\Rmnum{2}}$] is visible, thereby ensuring that star-formation has completely shut off. We did not do this for two key reasons: (i) the traditional method actively selects against galaxies that host narrow line AGN, which are more prevalent in post-starburst galaxies than galaxies with other star-formation histories, resulting in incomplete samples \citep{Wild et al. 2007, Wild et al. 2009, Yan et al. 2006}; (ii) the traditional method only selects old post-starburst galaxies, as starbursts themselves are not instantaneous, but rather have a decay time of $\sim0.3$ Gyr \citep{Barton et al. 2000, Freedman Woods et al. 2010, Hopkins Hernquist 2010, Wild et al. 2010}.
For further details of the sample selection, see \citet{Wild et al. 2010} and \citet{Rowlands et al. 2015}.

Prior to the analysis, the final sample was reduced to 335 galaxies. This was done after visual inspection of the galaxy images (SDSS $r$-band) which revealed that in case of 65 galaxies, the presence of bright field stars in the images could contaminate the measurements of the structural parameters. As a result, those galaxies were discarded. This does not affect the statistical property of our sample as the discarded galaxies were distributed evenly throughout all starburst age bins. 

\subsection{The control sample of star-forming and quiescent galaxies}\label{sec:rs_sample}

In order to relate the structure and morphology of galaxies evolving through the SB/PSB phase to that of both star-forming and passively evolving galaxies, we also selected control samples of 49 and 53 galaxies from the blue-cloud and red-sequence regions of the PC1-PC2 spectral index space, respectively. For that purpose, we used the same bulge-dominated base sample as during the selection of the SB/PSB galaxies. 
The control samples contain random selections of galaxies from the regions centred on the most populated parts of the blue cloud and the red sequence (marked by the blue/red box in Figure \ref{fig:vivplot}).
To eliminate potential biases due to varying galaxy structure with the stellar mass within both the blue cloud and the red sequence, we ensured that the stellar mass distributions of the galaxies within our control samples match that of the oldest (0.5 Gyr$<t_{SB}<$0.6 Gyr) SB/PSB galaxies in our main galaxy sample.


\section{Methodology}\label{sec:method}

To characterise the structure of the galaxies we applied a range of automated structural measures to the sky-subtracted $r$-band images of the (post-)starburst galaxies. We first defined a binary `detection mask' containing the pixels to be included in the structure measurements (Section \ref{sec:method_mask}). Then, we computed the standard measures of galaxy structure used in the literature (Section \ref{sec:method_oldparams}) as well a new measure of morphology introduced in this work, designed for detecting asymmetric morphological features characteristic of post-mergers (Section \ref{sec:method_ashape}).
\begin{figure*}
  \centering  
  \includegraphics[scale=0.85]{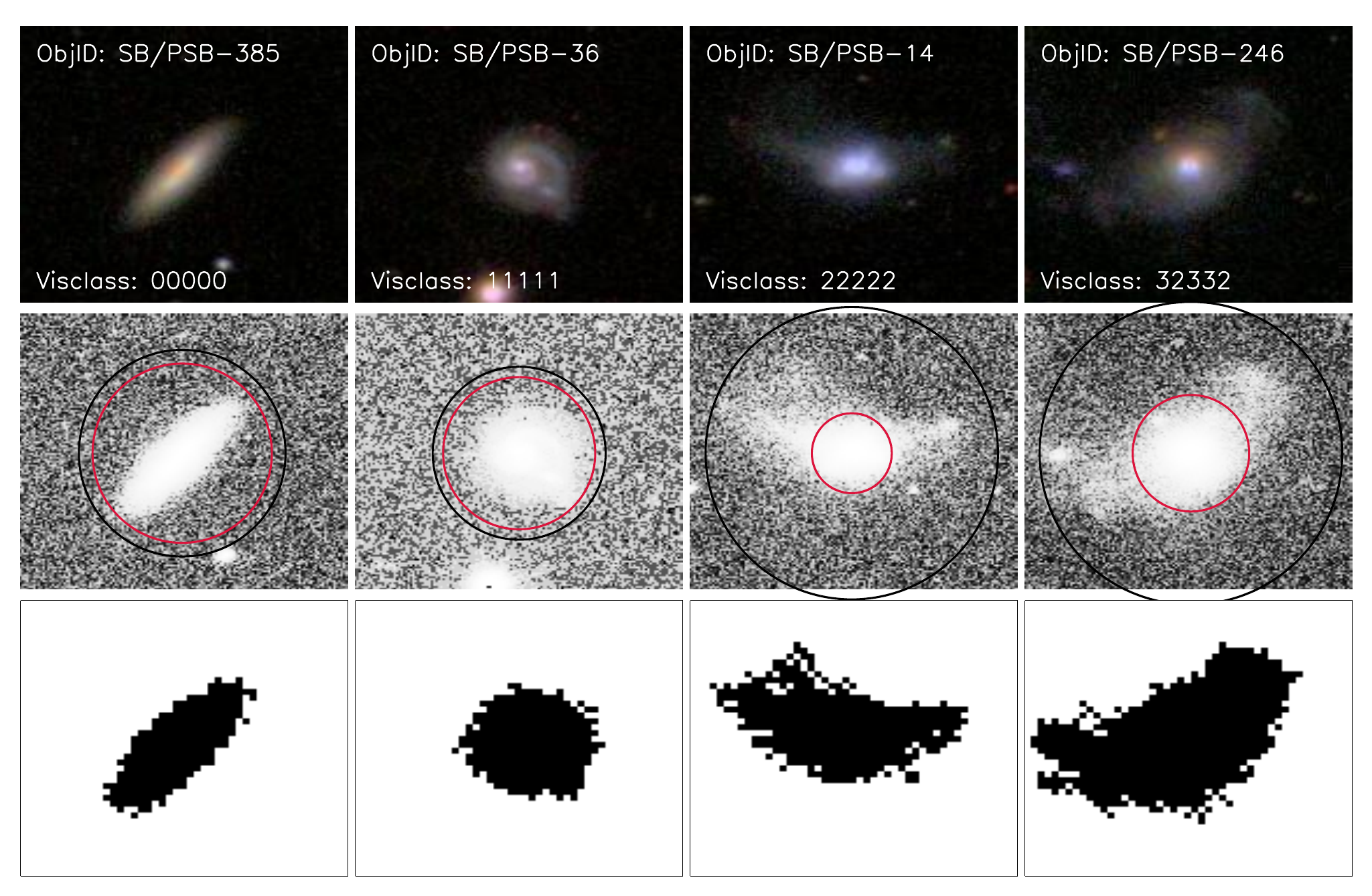}
  \caption{Examples of galaxies in the (post-)starburst sample with different levels of morphological disturbance (from left to right: regular morphology, low/moderate disturbance, major disturbance with elongated features, major disturbance with shell-like features). The horizontal panels show: \emph{top} - the false colour SDSS images with the corresponding visual classifications by 5 classifiers (described in detail in Section \ref{sec:resultspsb_visclass}); \emph{middle} - the SDSS $r$-band images with marked aperture sizes defined by the Petrosian radius, 2$R_{p}$ (red), and the radius introduced in Section \ref{sec:method_mask}, $R_{max}$ (black); \emph{bottom} - the corresponding binary detection masks (Section \ref{sec:method_mask}).}
  \label{fig:masksradii}
\end{figure*}
\subsection{Binary detection mask}\label{sec:method_mask}
To create the binary detection mask we employed an 8-connected structure detection algorithm, with 8-connected neighbours defined as all pixels touching one of the edges or corners of the central pixel. The algorithm searches for such neighbouring pixels with intensities above a threshold level defined as a function of the background sky level and its standard deviation. The algorithm is specifically designed to identify faint features in the outskirts of galaxies. To enhance the detectability of such features without significantly lowering the detection threshold and potentially dropping to the noise level, we passed the images through a $3 \times 3$ running average (mean) filter prior to the detection process. This reduced the noise in the images by not only amplifying the signal from the regions with low surface-brightness but also diminishing the pixellation effect in those regions where individual pixels are too faint to be detected. Then, starting from the central pixel as given by the SDSS position, the detection mask was built by accepting all pixels that are 8-connected to previously accepted pixels and have intensities above the chosen threshold. After some experimentation we found the optimal threshold for a robust detection of faint tidal features is 1 standard deviation above the sky background level. For the SDSS $r$-band images of the SB/PSB sample, this corresponds to a mean limiting surface brightness of 24.7 mag/arcsec$^{2}$.
In Figure \ref{fig:masksradii} we show examples of detection masks computed for SB/PSB galaxies with different levels of morphological disturbance, as classified by visual inspection of the SDSS $r$-band images (the exact criteria of the visual classification are explained in detail in Section \ref{sec:resultspsb_visclass}). In the case of ongoing mergers, with double nuclei joined by a bridge (Figure \ref{fig:morphsample1}, obj. 7, 9, 13, 17), the detection mask will include both components of the merging system.

The computation of the binary detection mask relies on a good estimate of the sky background level in the images. For a self-contained analysis, we estimated our own sky background levels rather than using SDSS values, although we find the two measurements agree well. We extracted pixels lying within a circular annulus with inner and outer radius of 20 and 40 times the full width half maximum (FWHM) of the light profile of each galaxy. This resulted in a large enough area and inner boundary sufficiently far from the central source to ensure a representative measurement of the sky background level \citep{Da Costa 1992, Simard et al. 2011}. We estimated the sky background level by the mode of the flux histogram, clipped iteratively at 3$\sigma$ \citep{Da Costa 1992, Berry Burnell 2000} until convergence of the mode (typically after a few iterations). We defined the mode as 2.5$\times$median - 1.5$\times$mean \citep{Bertin Arnouts 1996}.  

From the binary detection masks we estimated the galaxy radius, $R_{max}$, as the distance between the centre (brightest pixel within the mask) and the most distant pixel from that centre. We find that this definition of galaxy radius is an improvement for our purposes over the commonly used Petrosian radius, $R_{p}$ \citep{Petrosian 1976, Blanton et al. 2001, Yasuda et al. 2001}, in the case of galaxies with significant morphological disturbance. In the second row of Figure \ref{fig:masksradii} apertures at the new radii (black) are compared to those at twice the Petrosian radii (red). The latter value is typically used to recover the majority of flux for galaxies with regular morphologies \citep{Blanton et al. 2001}. The radii are similar for morphologically undisturbed galaxies but in the presence of significant morphological disturbance, $R_{max}$ includes the extended faint structures in the outskirts of galaxies, which can be excluded by $2R_{p}$.  Finally, we recomputed the sky background level within a circular annulus with inner and outer radius of one and two times $R_{max}$ \citep{Berry Burnell 2000}. This final estimate of the sky background level was subtracted from the images prior to the measurement of the morphological parameters (although it differs from the original one by no more than 3 digital counts per pixel, we note a drop in the standard deviation by as much as 50$\%$). 

\subsection{Structural parameters}\label{sec:method_oldparams}

Below, we list and briefly describe the measures of galaxy structure used in this study.
\\
\\
\emph{S{\'e}rsic index, n:}
For each galaxy, we fitted the S{\'e}rsic function \citep{Sersic 1963, Graham Driver 2005} to its surface brightness profile, extracted by considering the mean flux in pixel-thick concentric circular annuli defined by apertures centred at the brightest pixel of the galaxy image. 
\begin{equation}
I(R) = I_{e}\mbox{exp}\bigg\{-b_{n}\bigg[\bigg(\dfrac{R}{R_{e}}\bigg)^{1/n}-1 \bigg]\bigg\},
\label{eqn:sersic}
\end{equation}
where $I_{e}$ stands for the intensity at the galaxy's effective radius, $R_{e}$, enclosing half of the total galaxy light, and the constant $b_{n}$ is determined by the S{\'e}rsic index, $n$. During the fitting procedure, the S{\'e}rsic index was restricted to fall between 0.5 and 6.0.
\\
\\
\emph{Concentration index, C:}
We computed $C$ \citep{Kent 1985, Abraham et al. 1994, Bershady et al. 2000} from the growth curve radii $R_{20}$ and $R_{80}$, enclosing $20\%$ and $80\%$ of the galaxy's total light, respectively, calculated within a circular aperture defined by $R_{max}$:
\begin{equation}
C = 5\mbox{log}_{10} \bigg(\frac{R_{80}}{R_{20}}\bigg).
\label{eqn:c}
\end{equation}
\\
\emph{Asymmetry, A:}
 We followed the standard definition of rotational asymmetry \citep{Shade et al. 1995, Abraham et al. 1996, Conselice et al. 2000, Conselice 2003, Hernandez-Toledo et al. 2008}:
 \begin{equation}
A = \dfrac{\Sigma \mid I_{0}-I_{\theta} \mid}{2\Sigma \mid I_{0} \mid } - A_{bgr}
\label{eqn:a}
\end{equation}
where $I_{\theta}$ is the flux from an individual pixel in the galaxy image rotated $180^{o}$ about a chosen centroid (selected to minimise the values of $A$) and $I_{\theta}$ is flux from a pixel at the same position in the original, un-rotated image. The subtraction of $A_{bgr}$ (the asymmetry in the `background') accounts for the effect of noise on $A$.
We measured $A$ within a circular region defined by $R_{max}$ (rather than the commonly used $1.5 \times R_{p}$). We chose the rotation centroid out of a set of pixels accounting for the brightest $30\%$ of the galaxy's total light such as to minimise $A$:  for every pixel being treated as the rotation centroid, we computed $A$ and then selected the minimum value to represent our final measurement. This approach was taken to eliminate potential local asymmetry minima in the morphologically disturbed galaxies (\citealt{Conselice et al. 2000} found no such minima but their study did not specifically focus on galaxies with significant morphological disturbance, where this effect could become apparent). 
\\
\emph{Outer asymmetry, $A_{o}$:}
To enhance the signal from the low-surface-brightness regions in the outskirts of the galaxies we used a modified version of the asymmetry measure, which we call the `outer' asymmetry, $A_{o}$. It is computed similarly to the standard asymmetry parameter (Equation \ref{eqn:a}), but with an inner circular aperture containing the brightest $50\%$ of the total flux excluded from the measurement. A comparable definition of `outer' asymmetry was also recently developed by \citet{Wen et al. 2014}.
\begin{figure}
  \centering  
  \includegraphics[scale=0.85]{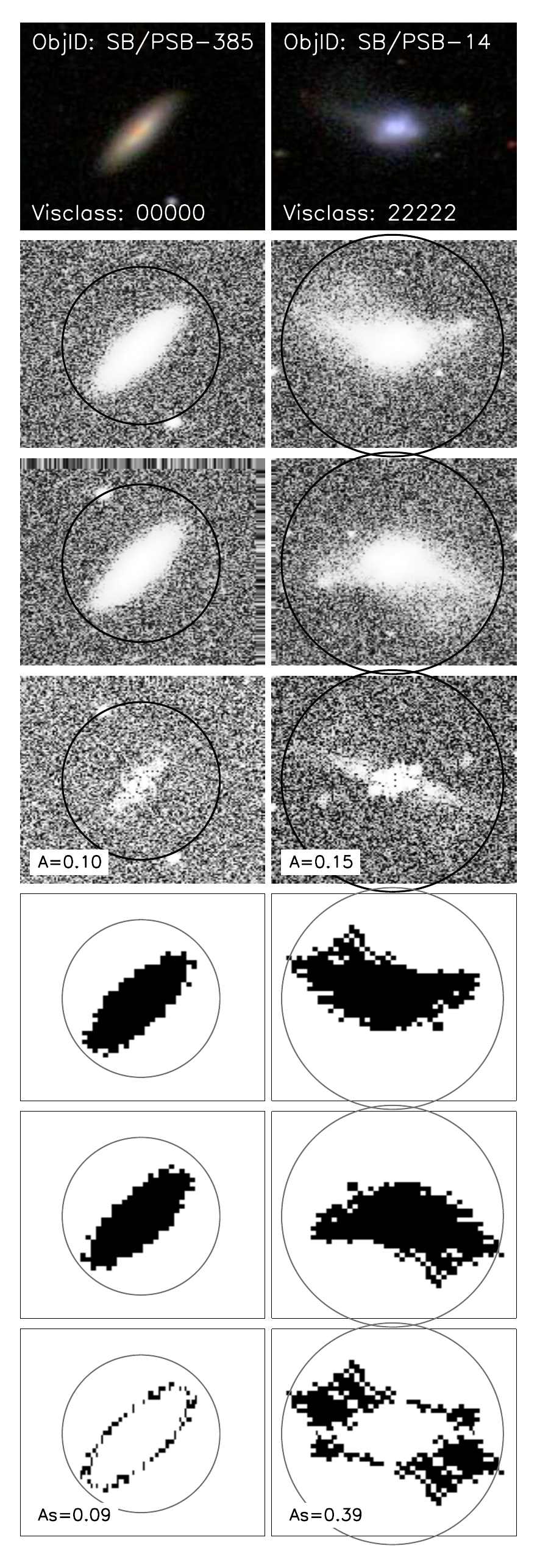}
  \caption{Measurement of $A$ and $A_{S}$ for a regular (left) and disturbed (right) galaxy. Rows 2, 3, 4 contain $r$-band images and rows 5, 6, 7, the detection masks: un-rotated, rotated and residual (absolute difference between the former two), respectively. The apertures show the extraction region defined by $R_{max}$.}
  \label{fig:masks}
\end{figure} \\
\\
\emph{Clumpiness, S:}
We measured $S$ following the standard definition \citep{Isserstedt Schindler 1986, Takamiya 1999, Conselice 2003}: 
\begin{equation}
S = \dfrac{\Sigma \mid I_{o} - I_{\sigma} \mid}{\Sigma \mid I_{o} \mid}
\label{eqn:s}
\end{equation}
 where $I_{\sigma}$ stands for pixel intensity in the smoothed galaxy image (using a Gaussian smoothing filter with width given by the radius containing 20\% of the total galaxy flux, $R_{20}$) and $I_{o}$ is that in the original image. We measured $S$ within an aperture defined by $R_{max}$ and centred on the highest intensity pixel. We excluded the central circular aperture at $R_{20}$ to eliminate the contribution from extremely bright centres present in some highly-concentrated galaxies (see e.g. \citet{Conselice 2003}).
\\
\\
\emph{Gini index, G:}
We used $G$ to measure the degree of inequality in the light distribution within images of the galaxies \citep{Glasser 1962, Abraham et al. 2003, Lotz et al. 2004}:
\begin{equation}
G = \dfrac{1}{2\overline{X}n(n-1)}\sum_{i}^{n}(2i-n-1)\mid X_{i}\mid
\label{eqn:g}
\end{equation}
with pixel intensities, $X_{i}$, in increasing order, $n$ - the total number of pixels assigned to the galaxy (in this work, using the detection mask), and $\overline{X}$ - the mean over all intensities.
$G$ is independent of the position of the galaxy's centre and is computed based on the rank-ordered cumulative distribution function of the pixel intensities within well specified boundaries. In this work, these boundaries were determined using the binary detection mask.
\\
\\
\emph{Moment of light, $M_{20}$:}
Following \citet{Lotz et al. 2004}, we computed the second order moment of the flux-weighted distance of the brightest pixels containing $20\%$ of the total flux ($M_{20}$) from the centre of the galaxy, to measure the spatial extent of the brightest galaxy regions:
\begin{equation}
\begin{split}
M_{20} = \mbox{log}_{10} \bigg( \dfrac{\sum_{i}M_{i}}{M_{tot}} \bigg), \mbox{while} \sum_{i}f_{i} < 0.2f_{tot} \\
M_{tot} = \sum_{i}^{n}M_{i} = \sum_{i}^{n}f_{i}[(x_{i}-x_{c})^{2}+(y_{i}-y_{c})^{2}]
\end{split}
\label{eqn:m}
\end{equation}
with the individual pixel coordinates denoted by $x_{i}$, $y_{i}$, and the centroid's coordinates given by $x_{c}$ and $y_{c}$.
We measured $M_{20}$ within boundaries defined by the binary detection masks, with respect to a free centroid parameter, computed by minimising the second order moment of the flux-weighted distance of all pixels, $M_{tot}$. In this work, the minimisation of $M_{tot}$ was performed using candidate centroid pixels within a region comprising the brightest $30\%$ of the galaxy's total flux.

\subsection{A new morphology measure: the `shape' asymmetry}\label{sec:method_ashape}
We introduce a new measure of asymmetry designed specifically to detect galaxies with low surface-brightness tidal features. We call the new measure the `shape' asymmetry ($A_{S}$). It is computed using the same mathematical expression as the standard asymmetry parameter (Equation \ref{eqn:a}); however the measurement is performed using the binary detection masks rather than the images of galaxies. This allows for equal weighting of all galaxy parts during the measurement, regardless of their relative brightness. The key difference between $A_{S}$ and the other asymmetry measures investigated in this work ($A$, $A_{o}$) is that the new parameter is purely a measure of morphological asymmetry and does not contain any information about the asymmetry of the light distribution within the galaxy. It is therefore not influenced by the presence of asymmetric bright galaxy components, like multiple nuclei, and it is highly sensitive to low surface-brightness regions of galaxies. This is illustrated in Figure \ref{fig:masks}.

As in the case of $A$, during the computation of $A_{S}$ the images were rotated around the flux-weighted minimum asymmetry centroid, to ensure that the asymmetry is measured with respect to the galaxy core rather than an arbitrary region selected by minimising the shape asymmetry itself. The measurement was performed within an extraction region defined by a circular aperture at $R_{max}$. The noise correction term ($A_{bgr}$, Equation \ref{eqn:a}) was omitted in the measurement of $A_{S}$: in contrast to the standard asymmetry measure, increasing the aperture size does not increase the amount of random noise as all `background' pixels within the detection mask are set to 0.

\section{Calibration and testing}\label{sec:results_test}
In this section we present the results obtained using the test sample for calibration of the structural parameters with our new binary detection masks (Section \ref{sec:resultstest_standard}) and testing of the newly introduced morphological measure (Section \ref{sec:resultstest_ashape}).

\begin{figure*}
  \centering  
  \includegraphics[scale=0.85]{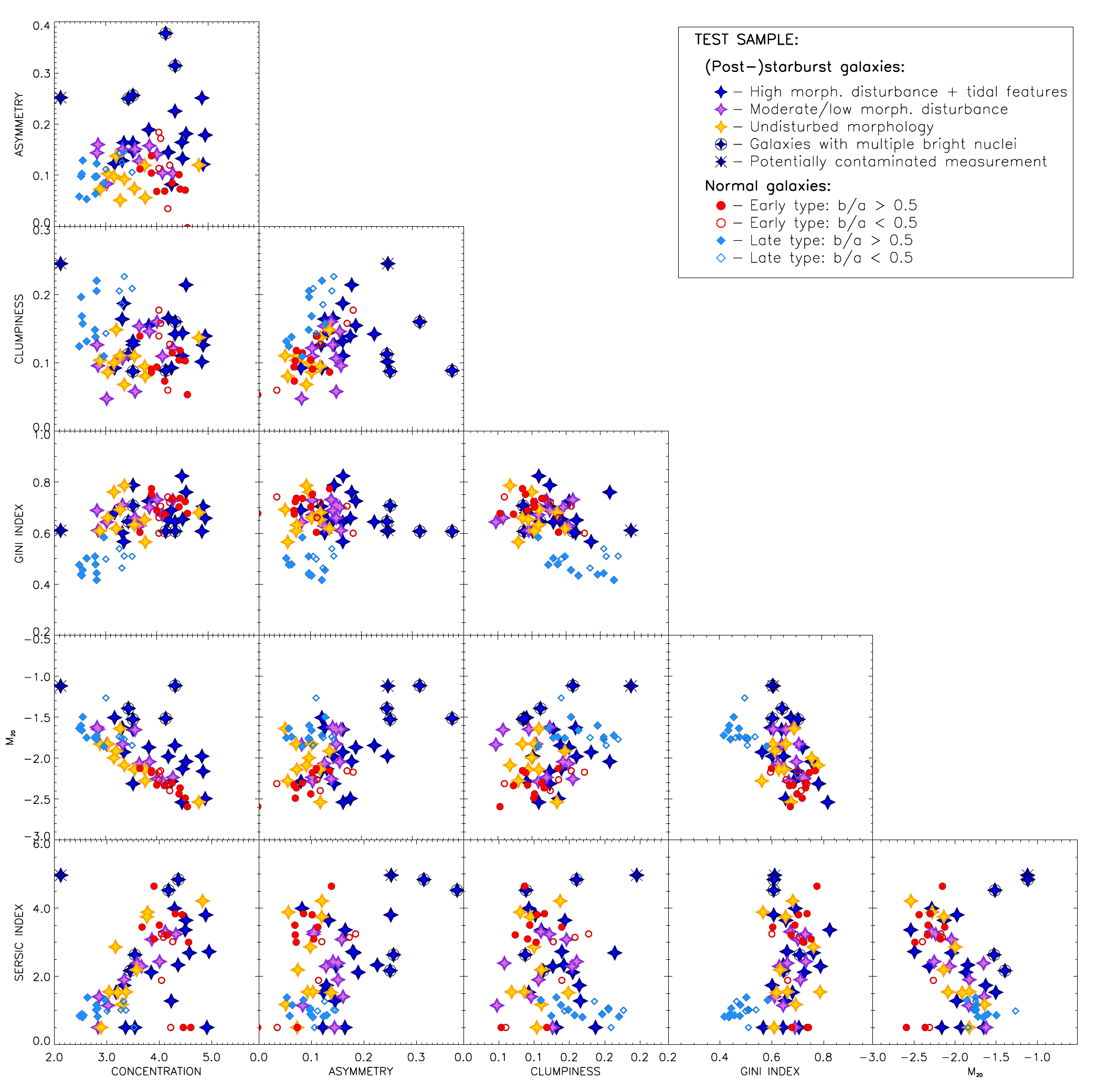}
  \caption{Relations between the six standard morphology measures (Sersic index [$n$], Concentration [$C$], Asymmetry [$A$], Clumpiness [$S$], Gini index [$G$], and $M_{20}$) computed for the test sample. The meaning of the symbols is described in the legend where galaxies are split into subsets based on visual classification and the SDSS \emph{fracdev} parameter (see Section \ref{sec:test_sample}).}
  \label{fig:test_all}
\end{figure*}

\subsection{The standard measures}\label{sec:resultstest_standard}
The early- and late-type galaxies (ETGs and LTGs, respectively) in the test sample provide a good standard for the measures of galaxy structure as they are expected to show values of $n$, $C$, $A$, $S$, $G$ and $M_{20}$ within well established ranges. Based on previous studies: 
\begin{itemize}
\item LTGs galaxies tend to have $n\sim1.0$, while ETGs show a range of values, typically with  $2.0\lesssim n \lesssim 4.0$ \citep{Sersic 1963}; 
\item $C$ ranges between 2.0 and 5.0, with highest values characteristic of ETGs ($C>4.0$) and decreasing to $C\sim3$ for LTGs \citep{Bershady et al. 2000, Conselice 2003, Hernandez-Toledo et al. 2008}; 
\item all normal galaxies tend to have low $A$, with values around 0.1 for ETGs, increasing towards later types to around 0.2 \citep{Conselice et al. 2000, Conselice 2003, Hernandez-Toledo et al. 2008}; 
\item $S$ also tends to increase towards later types, ranging from around 0.1 to 0.3 \citep{Conselice 2003, Hernandez-Toledo et al. 2008}; 
\item there is a clear separation in the values of $G$ found for ETGs and LTGs, with $G\sim0.6$ for the former and $G\sim0.4$ for the latter \citep{Abraham et al. 2003, Lotz et al. 2004}; 
\item $M_{20}$ takes on values lower than -2.0 for ETGs galaxies, and increases to around -1.5 for LTGs; galaxies with double nuclei have $M_{20}\sim-1.0$ \citep{Lotz et al. 2004}.
\end{itemize}

Figure \ref{fig:test_all} shows the six standard structural parameters computed for all galaxies in the test sample. The values obtained for the early- and late-type galaxies (ETGs and LTGs, respectively) are in general agreement with previous studies. ETGs appear more concentrated ($2.0\lesssim n \lesssim 5.0$ and $3.5\lesssim C \lesssim 4.5$) and have smoother light distribution ($S\sim0.1$) than LTGs ($0.5\lesssim n \lesssim 2.0$, $2.5\lesssim C \lesssim 3.5$, $S\sim0.2$). Both galaxy types show high symmetry under a rotation about $180^{o}$ ($0.05\lesssim A \lesssim 0.2$). ETGs and LTGs are also well separated from one another in the $G-M_{20}$ space, with the former showing more unequal light distributions ($0.4\lesssim G \lesssim 0.6$) and the presence of a compact bright nucleus (M$_{20}<-2$), than the latter ($0.6\lesssim G \lesssim 0.8$, M$_{20}>-2$). The separation between early and late types in the $A-S$ parameter space is not as strong as expected from the results of previous studies. This is likely a consequence of the relatively regular appearance and lack of prominent spiral arms in the LTG subsample (Appendix \ref{app:test}), leading to low values of both $A$ and $S$. 

Within the test sample, the morphologically disturbed galaxies are not easily distinguishable from galaxies with regular morphologies, when using the six standard structural parameters. Exceptions are galaxies with multiple bright nuclei (marked by circles in Figure \ref{fig:test_all}). In those cases, care should be taken when interpreting the values of $n$ and $C$, measured within the $R_{max}$, as they correspond to the entire (still merging) system rather than its individual components.
For most parameters, the values span the whole range characteristic of both ETGs and LTGs, with a slight bias toward the former. The exceptions are $G$ and $A$. 
The values of $G$ found for the morphologically disturbed galaxies are more similar to those found for highly-concentrated ETGs. We note that this tendency for such high values of $G$ could be a result of the sample selection as all morphologically disturbed galaxies were selected from a parent sample of galaxies with central SB/PSB signatures. This is supported by the fact that the undisturbed SB/PSB galaxies also show high values of $G$. 
Additionally, in morphologically disturbed galaxies the values of $G$ could potentially be enhanced by the presence of the faint tidal features which tend to add to the inequality in the light distributions in those galaxies (see \citealt{Lotz et al. 2004}).However,the fact that equally high values of $G$ were found in the morphologically undisturbed SB/PSB suggests that this contribution of faint tidal features to the measured values of $G$ may not be substantial.

The difference between morphologically disturbed galaxies and galaxies with regular morphologies is most noticeable in the values of $A$. However, only 35$\%$ (7/13) of the galaxies with the highest morphological disturbance in the test sample are well separated from the galaxies with regular appearance. Visual inspection of these galaxies shows that those with highest A generally have bright double nuclei.
The values of $A$ measured for galaxies with low morphological disturbance are comparable with those obtained for galaxies with regular morphologies.

We conclude that, while the standard measures investigated in this work are suitable for the structural analysis of normal galaxies as well as for identifying galaxies with multiple nuclei, they are not as well-suited for identifying galaxies with faint tidal features observed in post-mergers. In the following section, we present results for our new modified asymmetry measure, designed specifically for identifying galaxies with such features.

\begin{figure}
    \centering
    {\includegraphics[width=0.9\columnwidth]{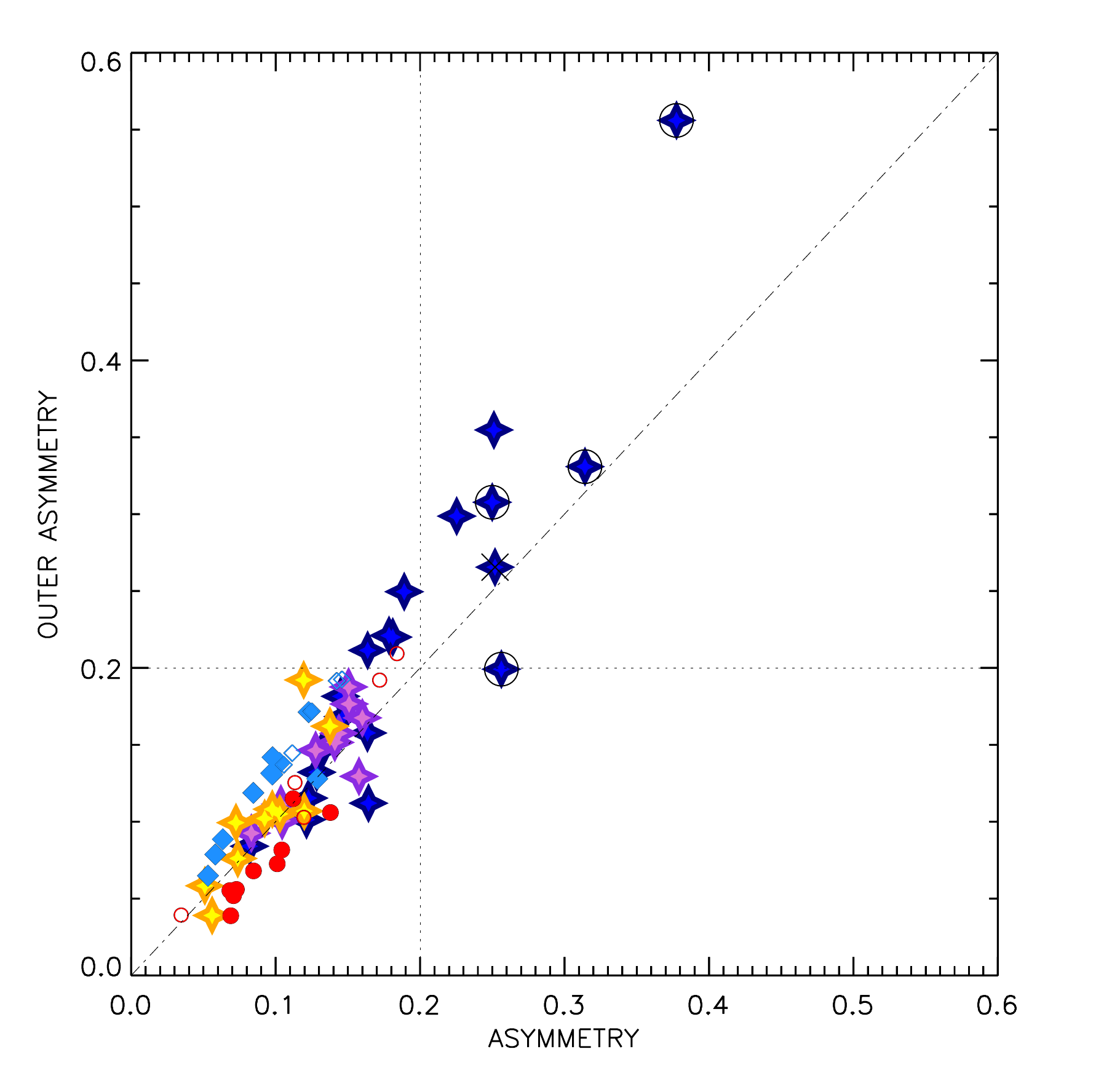}}\\
    {\includegraphics[width=0.9\columnwidth]{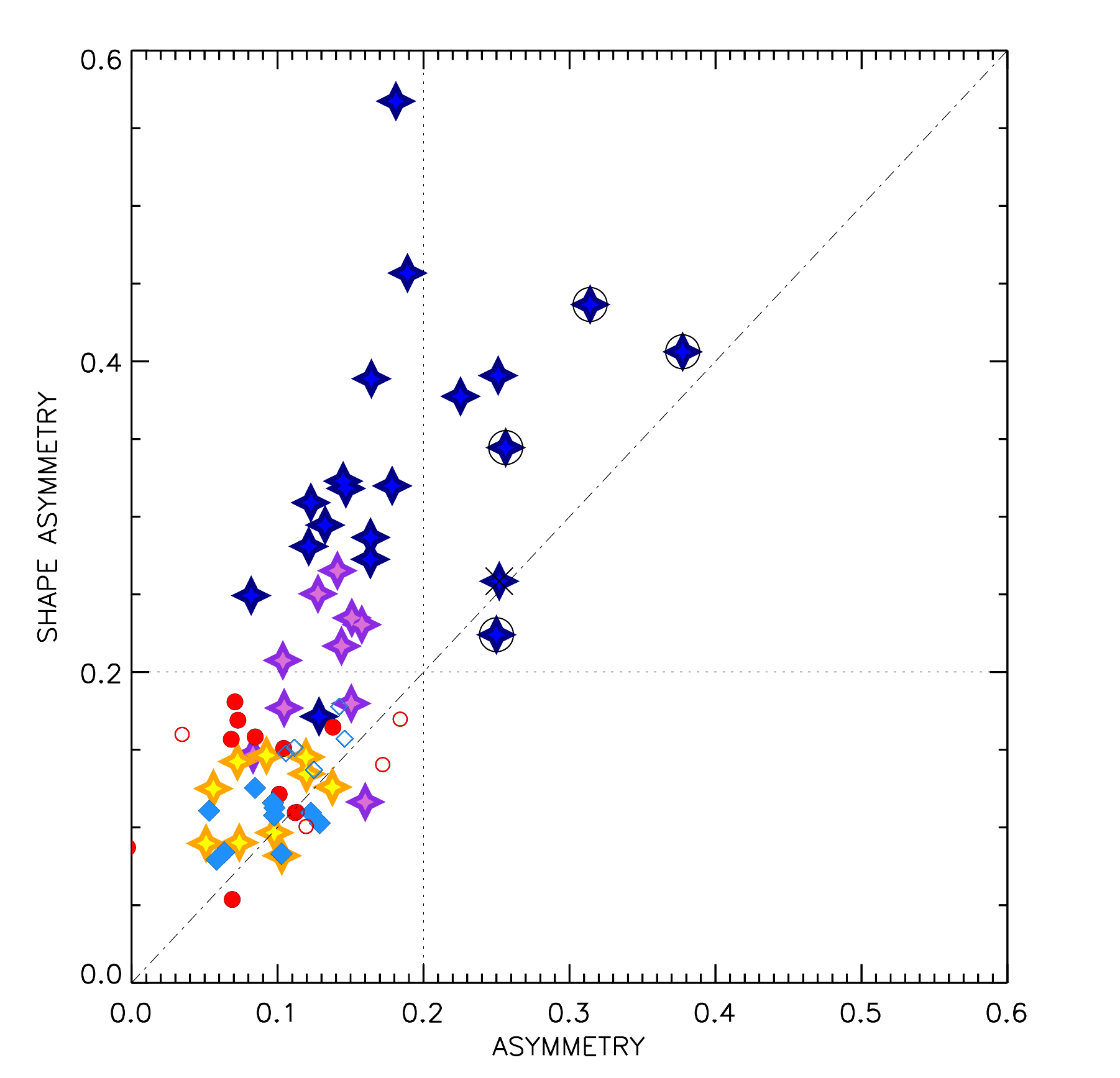}}\\
    {\includegraphics[width=0.9\columnwidth]{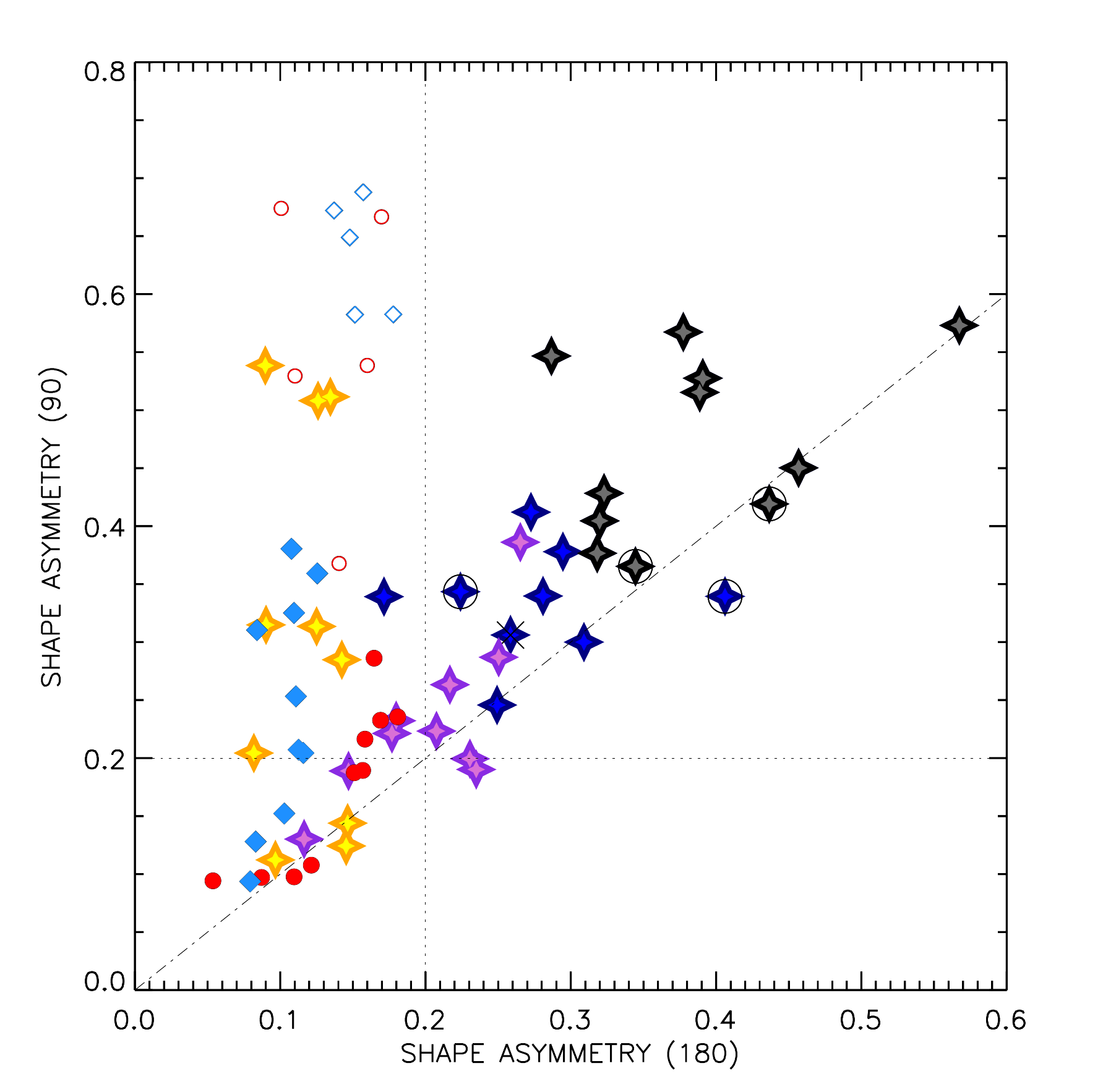}}
    \caption{For legend, see Figure \ref{fig:test_all}. In the bottom plot, the galaxies with significant morphological disturbance are further split into those with shell-like features (navy blue symbols) and elongated tails (black symbols).}
    \label{fig:A_morph}
\end{figure}

\subsection{The `shape' asymmetry}\label{sec:resultstest_ashape}

In Figure \ref{fig:A_morph} we compare the values of our new measure of morphology, $A_{S}$, with the two measures of structural asymmetry, $A$ and $A_{o}$. The dotted lines represent a threshold value at 0.2 used to separate between galaxies with regular morphologies and those with high morphological disturbance. The threshold was chosen empirically based on the fact that all galaxies with regular morphologies in the test sample have $A_{S}<0.2$ while $A_{S}\ge0.2$ is found only for galaxies with some level of morphological disturbance (detected by visual inspection). The situation is similar for $A$ and $A_{o}$, (with an exception of one elliptical galaxy having a slightly higher $A_{o}$). We compare the performance of the three measures of asymmetry by considering this threshold.

Figure \ref{fig:A_morph} shows that the shape asymmetry performs significantly better at separating galaxies with highest degree of morphological disturbance from those with regular appearance, than the other two asymmetry measures. Considering the above threshold value, $A_{S} \ge 0.2$ identifies $95\%$ of the galaxies with most prominent tidal features (dark blue symbols), compared with $35\%$ and $45\%$ picked out by $A$ and $A_{o}$, respectively. Visual inspection of the single galaxy omitted by $A_{S}$ revealed that the features present in the galaxy's structure form an azimuthally symmetric pattern, explaining the corresponding low value of $A_{S}$. We note that the new measure also recognises $60\%$ of galaxies with lower level of morphological disturbance (violet symbols), while both the standard asymmetry parameter and the outer asymmetry fail to pick out any such objects.

The ability of $A_{S}$ to detect galaxies with asymmetric tidal features comes from the independence of the measure of the flux distribution within the galaxies. In contrast, both $A$ and $A_{o}$ are flux-weighted measurements and therefore can become dominated by the brightest regions within galaxies, especially in the case of the former, where significant contribution to the measurement can come from multiple bright nuclei found in galaxies in pre-coalescence stages of a merger. Visual inspection identifies 4/7 ($57\%$) such galaxies in the test sample and all of them have $A\ge0.2$ (Figure \ref{fig:A_morph}). In the case of $A_{o}$, the contribution from multiple bright nuclei will not be as significant as in the case of $A$, as the inner aperture containing the brightest $50\%$ of the total light is omitted. The parameter should therefore be more sensitive to the low-surface brightness regions than $A$; however, as we show in Figure \ref{fig:A_morph}, it fails to identify more than half of the galaxies with highly disrupted morphologies. We also investigate outer asymmetries computed by cutting out inner apertures containing as much as 80$\%$ and 90$\%$ of the total flux, however, we find no significant improvement over $A_{o}$ with the inner $50\%$ cut out. This may stem from the differences in the light profiles of galaxies: while for some objects cutting out the inner circular aperture containing 50$\%$ of the total light may be optimal for separating its asymmetric outer features, it may prove insufficient or excessive for other galaxies. It is therefore not obvious how to define the inner circular aperture that would yield the most robust result. We also investigate an alternative approach to outer asymmetry by considering a central region with its geometry defined by the distribution of pixels accounting for a given fraction of the total galaxy light. However, we find that the irregularity of the inner region defined in such a way artificially increases the value of $A_{o}$. 

We conclude that the shape asymmetry is a much better indicator of the presence of asymmetric tidal features in galaxies than any of the standard structural parameters considered in this work, including the standard asymmetry parameter, $A$, and its modifications designed to measure the asymmetry of galaxy outskirts ($A_{o}$). We note that the shape asymmetry is purely a measure of galaxy morphology rather than structure. In order to gain information about the asymmetry of the light distribution within galaxies, the computation of the standard asymmetry parameter is a more suitable approach.

\subsubsection{Significance of rotation angles}
As shown in the study by \citet{Conselice et al. 2000}, while normal galaxies tend to show strong 180$^{o}$-symmetry, their projected shapes can introduce significant asymmetry under a $90^{o}$-rotation ($A_{90}$), which can serve as a good approximation for the directly measured minor-to-major axes ratio for statistical galaxy samples. In particular, galaxies seen `face-on' show $A_{90} \sim 0$, while those observed at higher inclination angles tend to reveal higher values, with $A_{90}$ reaching around 0.8 for the edge-on systems.

We applied the concept of asymmetry under a 90$^{o}$-rotation to the binary detection masks of galaxies within the test sample. As shown in Figure \ref{fig:A_morph}, spiral and elliptical galaxies are separated in $A_{S(90)}$ according to their axial ratio: galaxies with $b/a < 0.5$ tend to show higher asymmetries under a $90^{o}$ rotation. 
We find that $A_{S(90)}$ is roughly correlated with $A_{S(180)}$ for the morphologically disturbed galaxies, however the large scatter is caused by the variety of shapes of the tidal features. Within the test sample, galaxies with elongated tail-like features tend to have both higher $A_{S(180)}$ and $A_{S(90)}$, while those with shell-like features show lower $A_{S(90)}$.
Further tests of this aspect of the shape asymmetry will involve samples of galaxies with representative tidal features of different types (e.g. \citealt{Atkinson Abraham Ferguson 2013}), galaxies in earlier stages of a merger (e.g. \citealt{Ellison et al. 2013}), as well as simulated galaxy mergers \citep{Johansson Naab Burkert 2009}).

\begin{figure}
  \centering  
  \includegraphics[scale=0.95]{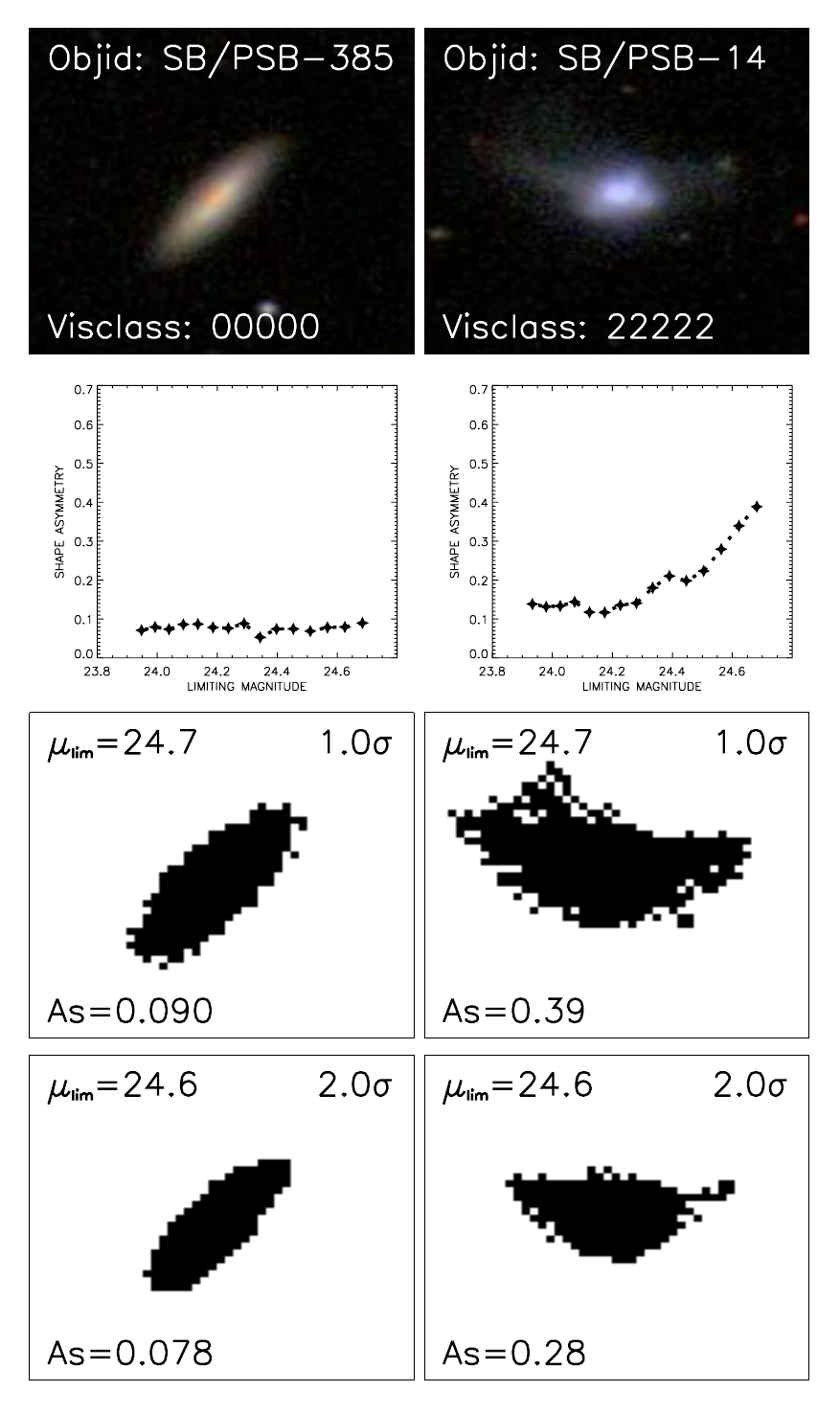}
  \caption{Dependence of the measurement of $A_{S}$ on the image quality (Section \ref{sec:imqual}) for a regular (left) and morphologically disturbed (right) galaxy (classified visually - see Section \ref{sec:resultspsb_visclass}). In the two bottom panels, the corresponding galaxy detection masks are presented, computed using threshold values equal to 1$\sigma$ and 2$\sigma$ above the sky background level estimate.}
  \label{fig:test_detThresh}
\end{figure}

\subsubsection{Image quality effects}\label{sec:imqual}
As the binary detection masks are created by assembling connected pixels above a specific threshold, the measurement of $A_{S}$ will depend on the choice of that threshold. Therefore, the limiting surface brightness of the images used to create the detection masks will be the measurement's main limitation.

For the SDSS $r$-band images, the optimal detection threshold was found to be $1\sigma$ above the sky background level, corresponding to $\mu_{lim}$= 24.7 mag/arcsec$^{2}$. Figure \ref{fig:test_detThresh} shows example results of tests of the dependence of $A_{S}$ on the limiting surface brightness of the galaxy images by increasing the detection threshold. Upon inspection of the resulting binary masks and the corresponding values of the shape asymmetry we find that, unsurprisingly, the behaviour of $A_{S}$ with limiting magnitude depends on the geometry of the features in the outskirts of the galaxies, as well as on their brightness relative to the central regions. For objects showing significant morphological disturbance, $A_{S}$ changes substantially and in general follows a decreasing trend with decreasing $\mu_{lim}$ (i.e. with decreasing image depth). Conversely, in the case of galaxies with regular morphologies, this effect is much less pronounced and the measurement of $A_{S}$ shows stability against varying $\mu_{lim}$. 

To summarise, $A_{S}$ provides a robust way of quantifying galaxy morphology and it is the most successful automated method to date at distinguishing between galaxies with asymmetric tidal features and those with regular morphologies. To accurately measure the shape asymmetry of galaxies, images of sufficient depth are required. The SDSS data used in this study are sufficiently deep in surface brightness to reveal faint structures in the outskirts of galaxies; however, using images of higher limiting magnitude could disclose further even fainter features in their morphology, consequently influencing their measured values of $A_{S}$. Therefore, great care must be taken when comparing measurements of $A_{S}$ between different surveys, and between galaxies with different redshifts.

\section{Results}\label{sec:resultspsb}

In this section we present the results of the structural and morphological analysis of the (post-)starburst galaxies. First, we summarise the results of a visual inspection of the galaxy images (Section \ref{sec:resultspsb_visclass}), and then we present the main findings of the automated approach (Section \ref{sec:resultspsb_params}). 

\subsection{Qualitative description}\label{sec:resultspsb_visclass}

We examined the $r$-band images of the 335 SB/PSB galaxies for features suggesting a past merger event, with an aim to determine the dependance of the presence of such features on the starburst age. The visual inspection was carried out independently by five reviewers and involved examining images of the SB/PSB galaxies as well as an additional sample of 300 continuously star-forming galaxies. The star-forming galaxies were selected from the blue cloud, using the same spectral indices as the SB/PSB galaxies and within the same redshift and stellar mass range (in this case, blue-cloud galaxies were more appropriate than red-sequence, because, due to their star formation activity, they are more likely to show disturbance in their structure than galaxies in which the star formation has already quenched). All images were randomised prior to inspection. Each galaxy was then classified, according to a pre-agreed scheme, as an object with:
\begin{itemize}
\item regular morphology, no signs of disturbance pointing to any kind of interaction (class 0), 
\item low/moderate level of morphological disturbance suggesting a possible interaction (class 1), 
\item high morphological disturbance (obvious post-merger) with elongated tidal features, e.g. tails (class 2), 
\item high morphological disturbance (obvious post-merger) with shell-like features (class 3). 
\end{itemize}
Examples of the different classes of galaxies are presented in the top row of Figure \ref{fig:masksradii}, with the numeric classes assigned by the 5 classifiers. The visual inspection has shown that the galaxies in our SB/PSB sample have a variety of morphologies. Many, in particular those of young starburst ages, show signs of a past merger, including disturbed morphologies and presence of tidal features. Ongoing mergers with double bright nublei constitute only $\sim3\%$ of the whole sample.

\begin{figure}
  \centering  
  \includegraphics[scale=0.5]{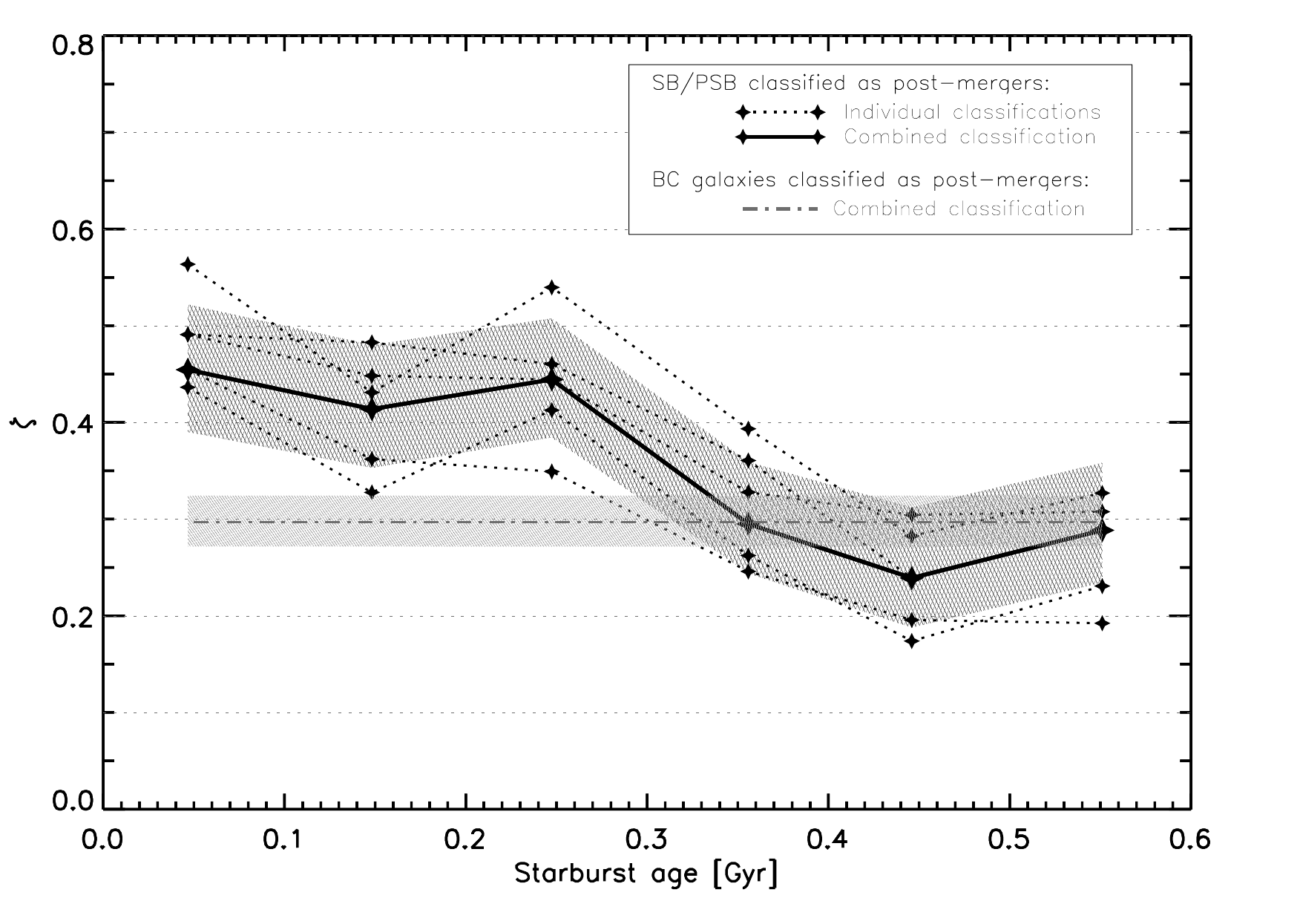}
  \caption{Visual classification of the 335 SB/PSB galaxies and a control sample of 300 star-forming galaxies. $\zeta$ represents the fraction of SB/PSB galaxies with morphological disturbance per starburst age bin (solid line), where at least three reviewers agreed on the presence of morphological disturbance. The dotted lines correspond to the classifications by individual reviewers and the dashed line shows the fraction of normal star-forming galaxies with morphological disturbance. The starburst age does not apply to the control sample. The shaded regions represent the binomial $1\sigma$ confidence intervals, calculated after \citet{Cameron 2010}}
  \label{fig:visual}
\end{figure}

In Figure \ref{fig:visual} we define $\zeta$ as the fraction of SB/PSB galaxies showing signatures of a past (or ongoing) merger (disturbance in morphology, presence of tidal features such as tails, arms or shells - i.e. classes 1, 2, 3), calculated independently in different age bins with a width of 0.1 Gyr. As visual classification is subjective, the individual classifications were not always in complete agreement. For a compromise of reliability and detectability, we discuss the `combined classification', in which case $\zeta$ includes only those galaxies for which at least three reviewers agreed on the presence of morphological disturbance. As shown in Figure \ref{fig:visual}, about $45\%$ of the youngest SB/PSB galaxies (t$_{SB}$ $<$ 0.1 Gyr) in the sample show features characteristic of a post-merger. Furthermore, there is a decreasing trend in $\zeta$ with starburst age. For the older PSB galaxies in the sample (t$_{SB}$ $>$ 0.3 Gyr), we find $\zeta\sim30\%$, which is consistent with the fraction of morphological disturbance found in our control sample of ordinary blue-sequence galaxies.

\subsection{Quantitative measures of structure and morphology}\label{sec:resultspsb_params}

\begin{figure*}
  \centering  
  \includegraphics[scale=0.77]{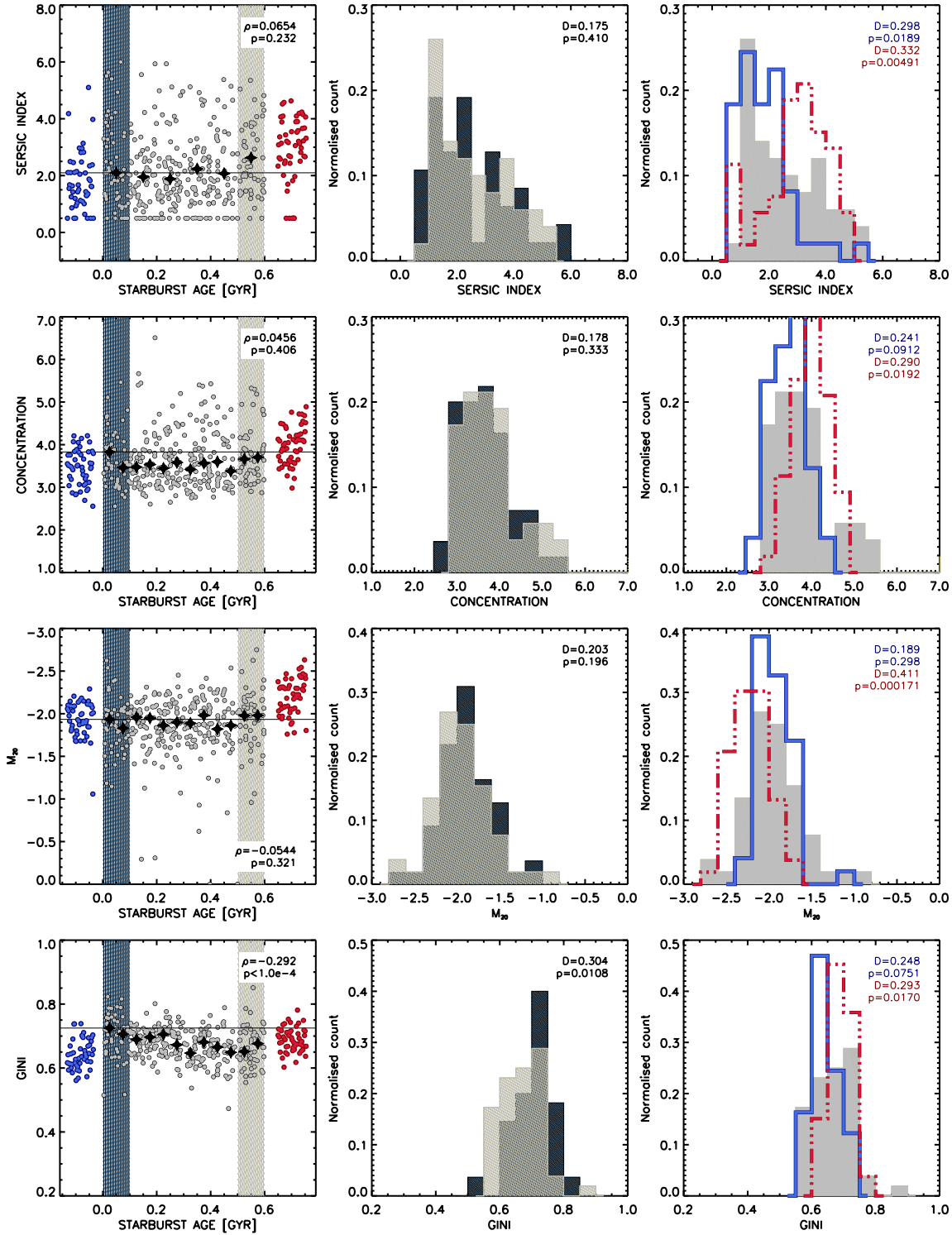}
  \caption{\emph{Left column}: the evolution of the S{\'e}rsic index ($n$), concentration index ($C$), moment of light ($M_{20}$) and the Gini index ($G$) as a function of the starburst age for the (post-)starburst sample (grey symbols - values for individual galaxies, black symbols - median values per age-bin). The horizontal lines denote the initial (median) level for each parameter. The blue/red data points correspond to the control samples of blue-cloud/red-sequence galaxies which are placed on the $x$-axis at a range of arbitrary starburst ages for clarity. The shaded regions mark the young and old subsets of (post-)starburst galaxies. The numbers quoted are: the Spearman rank correlation coefficient ($\rho$) and the two-sided significance of the coefficient's deviation from zero ($p$). \emph{Middle column}: histograms show the distributions of the parameter values for the young (dark grey) and old (light beige) (post-)starburst galaxies. \emph{Right column}: histograms correspond to the old post-starburst galaxies (light beige) and the blue-cloud/red-sequence galaxies (blue/red). $D$ is the K-S statistics and $p$, the corresponding probability of the null hypothesis.}
  \label{fig:nCGM20_age}
\end{figure*}

\begin{figure*}
  \centering  
  \includegraphics[scale=0.77]{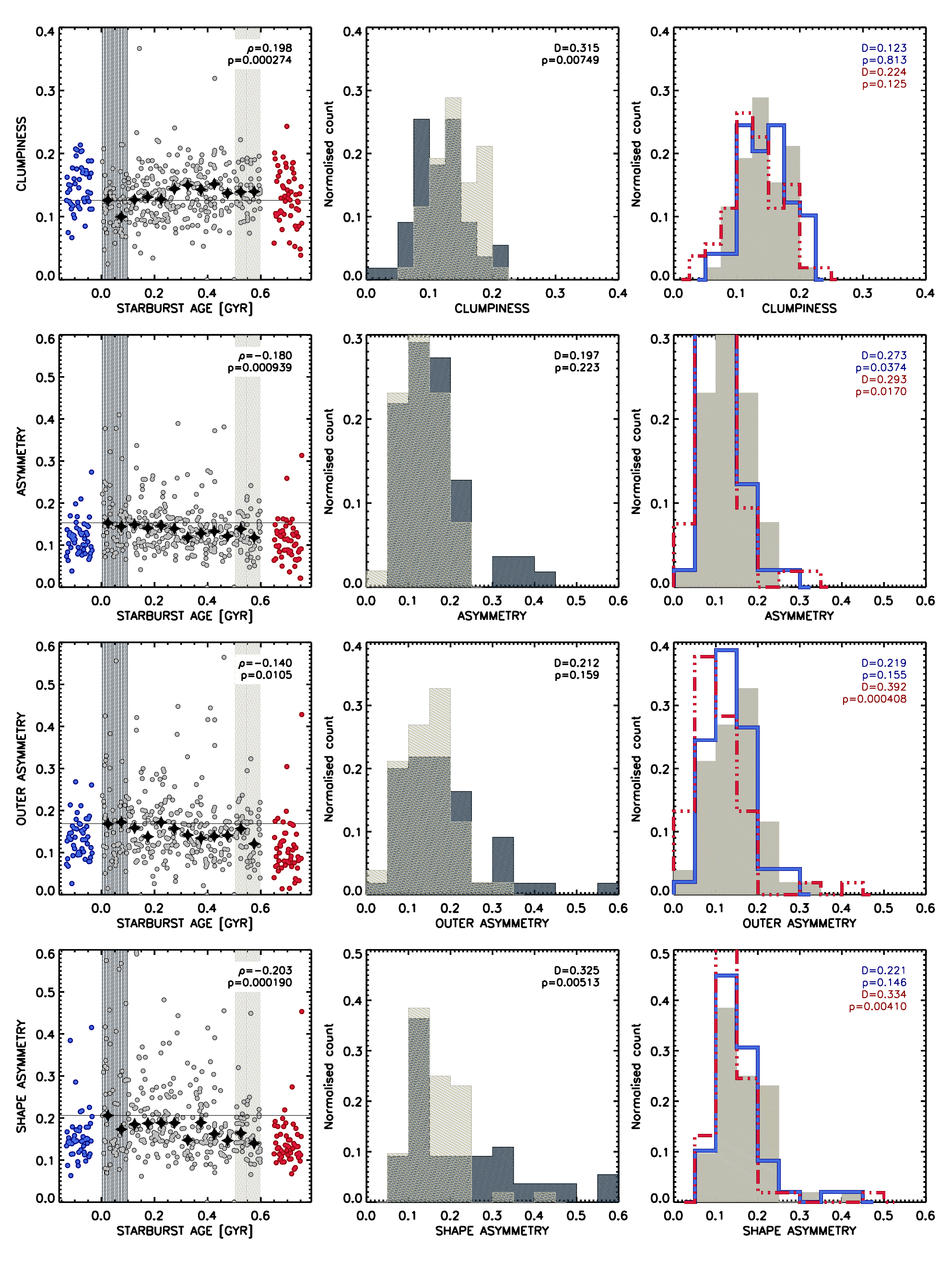}
  \caption{\emph{Left column}: the evolution of the measure of clumpiness ($S$), standard asymmetry ($A$), outer asymmetry ($A_{o}$) and shape asymmetry ($A_{S}$) as a function of the starburst age for the (post-)starburst sample (\emph{grey} - values for individual galaxies, \emph{black} -median values per age-bin). The horizontal lines denote the initial (median) level for each parameter. The blue/red data points correspond to the control samples of blue-cloud/red-sequence galaxies which are placed on the $x$-axis at a range of arbitrary starburst ages for clarity. The shaded regions mark the young and old subsets of (post-)starburst galaxies. The numbers quoted are: the Spearman rank correlation coefficient ($\rho$) and the two-sided significance of the coefficient's deviation from zero ($p$). \emph{Middle column}: histograms show the distributions of the parameter values for the young (dark grey) and old (light beige) (post-)starburst galaxies. \emph{Right column}: histograms correspond to the old post-starburst galaxies (light beige) and the blue-cloud/red-sequence galaxies (blue/red). $D$ is the K-S statistics and $p$, the corresponding probability of the null hypothesis.}
  \label{fig:SAAoAs_age}
\end{figure*}

In Figures \ref{fig:nCGM20_age} and \ref{fig:SAAoAs_age}, we present all measures of galaxy structure and morphology considered in this work, computed for the SB/PSB sample as a function of starburst age ($t_{SB}$), as well as for the control samples of star-forming blue-cloud and quiescent red-sequence galaxies (plotted with arbitrary ages for clarity).  For each parameter, we show individual values per galaxy and median values calculated in 0.1-Gyr wide age-bins (left column). We also quote the Spearman rank correlation coefficient ($\rho$) and the corresponding $p$-value (two-sided significance of the coefficient's deviation from zero). All values are summarised in Table \ref{tab:table1}, where we comment on the general characteristics of our sample, inferred from the individual parameters. Figures \ref{fig:nCGM20_age} and \ref{fig:SAAoAs_age} also show distributions of the parameter values: in the middle panel, for the youngest ($t_{SB}\le0.1$Gyr) and oldest (0.5 Gyr$\le t_{SB}\le $0.6 Gyr) (post-)starburst galaxies in the sample, and in the right panel, for the oldest (post-)starburst galaxies and the control sample of star-forming and quiescent galaxies. For quantitative comparison of the distributions, we performed a Kolmogorov-Smirnov (K-S) test. We quote the obtained values of the K-S statistics ($D$) and the corresponding probability of the null hypothesis ($p$) in the respective panels of Figure \ref{fig:nCGM20_age} and \ref{fig:SAAoAs_age} .

\begin{table}
 \caption{Structural and morphological evolution of SB/PSB galaxies (see \emph{left} panel of Figures \ref{fig:nCGM20_age} and  \ref{fig:SAAoAs_age}). First column: range of parameter values (for the majority of galaxies); second column: trend with the starburst age quantified by the Spearman rank correlation coefficient and the corresponding $p$-value; third column: comments on the mean sample characteristics. }
 \label{tab:table1}
 \begin{tabular}{ccl}
  \hline
Typical range & Trend & Comments \\
  & with $t_{SB}$ & (mean characteristics)  \\
& (Spearman) & \\
  \hline
  \hline
 & & - light profiles of \\
  $1.0 \lesssim n \lesssim 3.0$ & $\rho\sim0.06$ & early disks; \\
 & $p\sim0.23$  & - no significant trend\\
 & & with starburst age; \\
 \hline
 & & - moderate \\
 $2.9\lesssim C \lesssim4.2$ & $\rho\sim0.04$  & concentration;  \\
 & $p\sim0.41$  & - no significant trend \\
 & & with starburst age; \\
  \hline
 & & - moderately extended \\
 $-2.4\lesssim M_{20} \lesssim-1.4$ & $\rho\sim-0.05$ & single nucleus;  \\
  & $p\sim0.32$  & - no significant trend \\
 & & with starburst age; \\
  \hline
 & & - highly unequal light \\
 $0.55\lesssim G \lesssim0.8$ & $\rho\sim-0.29$ & distribution;  \\
  & $p<1.0e^{-4}$  & - moderate decline \\
 & & with starburst age; \\
  \hline
 & & - moderately `clumpy' \\
 $0.08\lesssim S \lesssim0.2$ & $\rho\sim0.20$ & light distribution;  \\
  & $p\sim1.0e^{-4}$  & - moderate increase \\
 & & with starburst age; \\
  \hline
 $0.05\lesssim A \lesssim0.25$ & $\rho\sim-0.18$ & - moderately declining\\
& $p\sim1.0e^{-3}$  & asymmetry  \\
  && with starburst age; \\
  \hline
  $0.05\lesssim A_{o} \lesssim0.25$ & $\rho\sim-0.14$ &  - moderately declining \\
 & $p\sim0.01$  & asymmetry  \\
  && with starburst age; \\
  \hline
  $0.05\lesssim A_{S} \lesssim0.35$ & $\rho\sim-0.20$ & - moderately declining \\
 & $p\sim1.0e^{-4}$ & asymmetry  \\
  && with starburst age; \\
  \hline
  \hline
 \end{tabular}
\end{table}

\begin{table}
 \caption{Age evolution of the fraction of SB/PSB galaxies with strong asymmetric features (outliers in the plots of $A$, $A_{o}$ or $A_{S}$, see Figure \ref{fig:SAAoAs_age}) . The percentages in the second and third column correspond to the fraction of SB/PSB galaxies with $t_{SB}<0.3$ and $t_{SB}>0.3$, respectively, with values of either $A$, $A_{o}$ or $A_{S}$ greater than 0.2. The last column shows the trends with the starburst age quantified by the Spearman rank correlation coefficient and the corresponding $p$-value.}
 \label{tab:table2}
 \begin{tabular}{cccc}
  \hline
Asymmetry & Outliers with & Outliers with & Trend in \\
measure & $t_{SB}<0.3$ & $t_{SB}>0.3$  & the fraction  \\
& & & of outliers \\
\hline
\hline
$A$ & $16\%$ & $8\%$ & $\rho\sim -0.71$\\
&&& $p\sim0.11$ \\
\hline
$A_{o}$ & $27\%$ & $14\%$ &$\rho\sim -0.77$ \\
&&& $p\sim0.07$ \\
\hline
$A_{S}$ & $43\%$ & $21\%$ & $\rho\sim -0.94$\\
&&& $p\sim10^{-3}$ \\
 \hline
  \hline
 \end{tabular}
\end{table}

Considering the measures pertaining to the central concentration of light and the spatial extent of the galaxy nucleus, we find that most galaxies in the SB/PSB sample show characteristics of early-type disks:
$1.0\lesssim n\lesssim 3.0$, $3.0\lesssim C\lesssim 4.0$, $M_{20}$ peaking around -2.0 (see e.g. \citealt{Bershady et al. 2000, Conselice 2003, Lotz et al. 2004, Hernandez-Toledo et al. 2008}). As suggested by the low values of the Spearman rank correlation coefficient, there is no significant trend in either $n$, $C$ or $M_{20}$ with starburst age (see left panel of Figure \ref{fig:nCGM20_age}). This points to little evolution of the central light concentration over the first 0.6 Gyr after the starburst. Direct comparison of the young and old (post-)starburst galaxies shows little difference in the distributions of all three parameters (middle panel, Figure \ref{fig:nCGM20_age}) and the K-S test suggests that they are 
likely to have been drawn from the same distribution. Comparison of the old subset with the control sample of red-sequence galaxies reveals that, after 0.6 Gyr since the last starburst, the (post-)starburst galaxies are not as highly concentrated as the galaxies populating the red sequence; rather, their structure ($n$, $C$, $M_{20}$) bears closer resemblance to star-forming blue-cloud galaxies (right panel, Figure \ref{fig:nCGM20_age}).

The values of the Gini index found for the youngest galaxies in the (post-)starburst sample are as high as those obtained for the red-sequence galaxies (right panel, Figure \ref{fig:nCGM20_age}). Contrary to $n$, $C$ and $M_{20}$, $G$ is a measure independent of the position of the galaxy centre or assumptions of azimuthally symmetric apertures, and it is designed to pick out changes in the light distribution on a pixel scale. It should therefore be more sensitive to subtle variations in the light distribution of galaxies following a starburst than the other measures. Consequently, high $G$ does not necessarily imply high central concentration.
$G$ shows the most pronounced correlation with the starburst age out of all measures considered in this work ($\rho=-0.3$, $p<1.0^{-4}$). The decreasing tendency suggests that SB/PSB galaxies tend to have more uniform light distribution as they age, which could be a consequence of the decaying starburst as well as, to some extent, diminishing prominence of the faint tidal features. We do not attribute this effect to contribution from multiple nuclei as their presence is indicated only in a few galaxies by $M_{20}$ (the dearth of objects with $M_{20}\ge-1.0$). This was confirmed during the visual inspection of the galaxy images (only $\sim3\%$ classified as ongoing mergers).
The fading central starburst could also be the reason for the increasing tendency in clumpiness as a function of starburst age (left panel, Figure \ref{fig:SAAoAs_age}). As the amount of light from the star-bursting region decreases, the less luminous off-centre star-forming regions are being uncovered, increasing the resulting values of $S$ (although the parameter is designed to exclude the bright central regions from the measurement, the size of the excluded central aperture will not always coincide with that of the star-bursting region).

All three measures of rotational asymmetry, $A$, $A_{o}$ and $A_{S}$, show a decreasing tendency with the starburst age (left column, Figure \ref{fig:SAAoAs_age}) but the trend is most pronounced in the values of the shape asymmetry ($\rho=-0.2$, $p\sim10^{-4}$). We also find an excess of young galaxies with high $A_{S}$ when comparing the distributions of $A_{S}$ for the young and old subsets (middle panel, Figure \ref{fig:SAAoAs_age}) of the (post-)starburst sample. In Table \ref{tab:table2}, we compare the number of such `outliers' and its evolution with starburst age for all three measures of asymmetry, using the threshold value of 0.2 defined in Section \ref{sec:resultstest_ashape}. We find the highest number of outliers for the shape asymmetry: 43$\%$ of galaxies with $t_{SB}<0.3$ Gyr show $A_{S}\ge0.2$, and the fraction drops to $21\%$ for galaxies with $t_{SB}>0.3$ Gyr. In the last column of Table \ref{tab:table2}, we quantify the trend in the fraction of galaxies with $A$, $A_{o}$ and $A_{S}$ greater than 0.2 as a function of the starburst age (in 0.1 Gyr-wide age bins), by means of the Spearman rank correlation coefficient and the corresponding $p$-value. We find a decreasing tendency for all three measures, but a particularly strong and statistically significant trend for the shape asymmetry ($\rho\sim-0.9$, $p\sim10^{-3}$). 
This declining number of the outliers suggests that, as (post-)starburst galaxies age, they are less likely to show the presence of asymmetric features in their structure. In the case of the standard asymmetry measure, the observed decline is likely to be associated with changes in the symmetry of the central bright galaxy regions. The decrease in the number of galaxies with high shape asymmetry, as a function of the starburst age, points directly to morphological transformations and implies that (post-)starburst galaxies are less likely to show the presence of tidal features, characteristic of post-mergers, as they age.

To summarise, the quantitative analysis of the structure of local (post-)starburst galaxies with high stellar mass surface density, reveals moderate central concentration, light profiles and spatial extent of the brightest regions characteristic of early-type disk galaxies, as expected for the sample selection. We do not find any significant evolution of these properties during the first 0.6 Gyr following the most recent starburst.
The distribution of light within the (post-)starburst galaxies shows a high level of inequality, as measured by the Gini index, (particularly for those with youngest starbursts), and this inequality decreases with the starburst age. This could be an effect of the decaying starburst and/or of the fading faint tidal features in the outskirts of those galaxies, however, the latter is not expected to be significant, as discussed in Section \ref{sec:resultstest_standard}. Using the shape asymmetry we find that, within the first 0.6 Gyr after the most recent starburst, (post-)starburst galaxies are less likely to show the presence of tidal features characteristic of post-mergers as they age. 

\begin{figure*}
\centering
  \includegraphics[width=0.97\columnwidth]{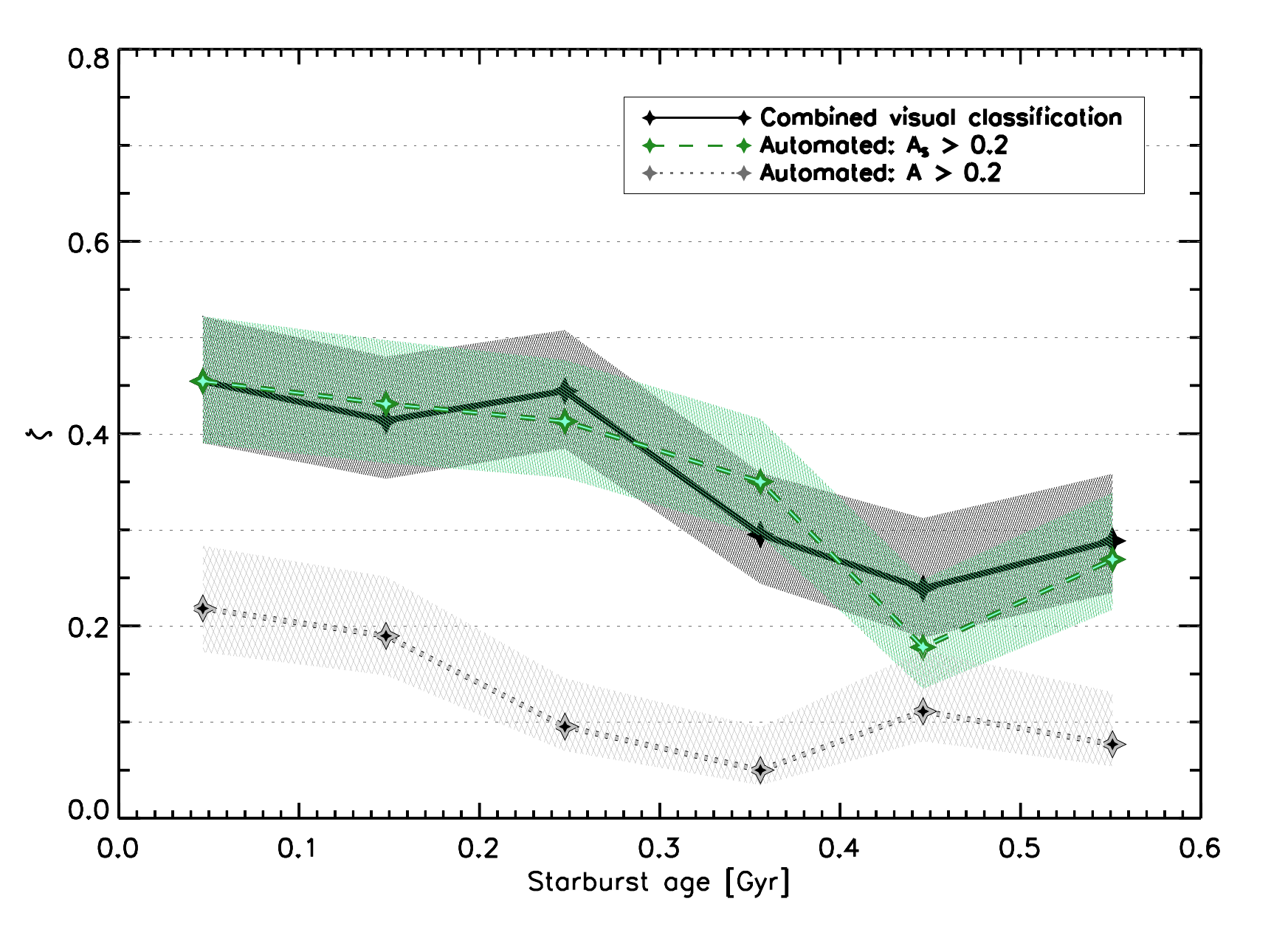}
  \includegraphics[width=\columnwidth]{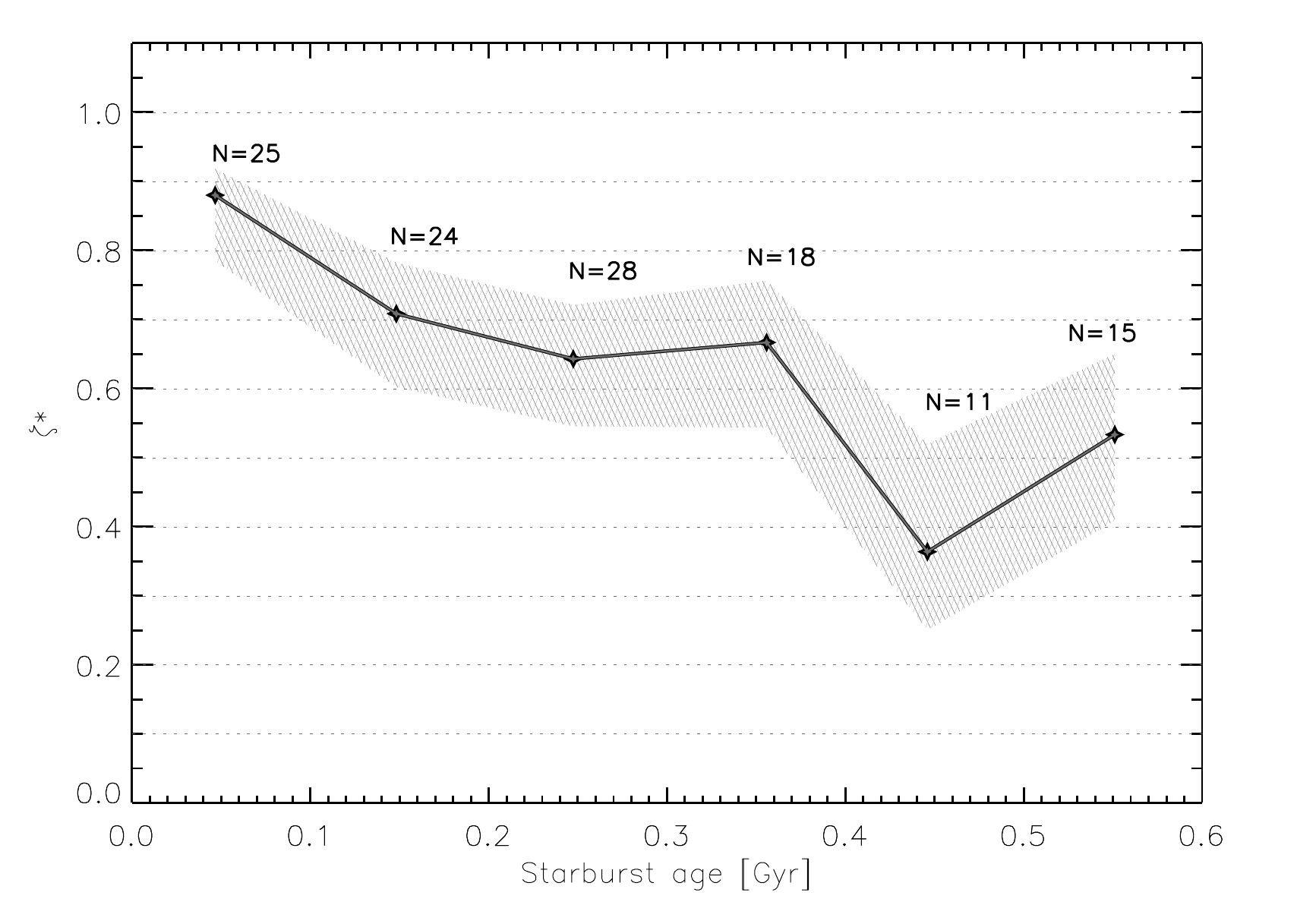}
\caption{\emph{Left panel:} The fraction of post-starburst galaxies showing signatures of a past (or still ongoing) merger ($\zeta$) computed automatically by considering galaxies with $A \ge 0.2$ (dashed green line) and $A_{S} \ge 0.2$ (dotted grey line). For comparison, we also show the combined visual classification presented in Figure \ref{fig:visual} (solid black line).  \emph{Right panel:} The fraction of post-starburst galaxies classified as past/ongoing merger candidates (by minimum three reviewers), for which $ A_{S} \ge 0.2$. In both plots, the shaded regions show the binomial $1\sigma$ confidence intervals, calculated after \citet{Cameron 2010}. }
\label{fig:visauto}
\end{figure*}

\section{Discussion}

Both visual and automated analyses of the (post-)starburst galaxies point to a declining number of galaxies with high morphological disturbance as a function of  starburst age. Below, we present a quantitative comparison of the visual vs. automated approaches used in this paper (Section \ref{sec:discussion_visauto}). We then discuss the findings of this study in the context of galaxy evolution (Section \ref{sec:discussion_galevo}).

\subsection{Detecting tidal features - visual versus automated approach}\label{sec:discussion_visauto}
In the left panel of Figure \ref{fig:visauto} we compare the qualitative analysis of the (post-)starburst galaxies based on visual classification, with both the standard and new automated measures of asymmetry. As in Section \ref{sec:resultspsb_visclass}, we define $\zeta$ as the fraction of (post-)starburst galaxies classified as post-merger candidates in a given age bin. In the case of the visual approach, we show the combined classification, including only those galaxies for which at least three reviewers agreed on the presence of morphological disturbance. For the automated classification, we consider the threshold value introduced in Section \ref{sec:resultstest_ashape} where an automatically selected post-merger candidate has $A_{S}\ge0.2$. The automated classification computed using the shape asymmetry reproduces almost exactly the result of the human visual classification, pointing to a declining trend in $\zeta$ with the starburst age. 

In the right panel of Figure \ref{fig:visauto}, we define $\zeta^{*}$ - the fraction of the \emph{visually identified post-merger candidates} which have $A_{S}\ge0.2$, calculated in a given age bin. Above each data point, we quote the total number of galaxies (both with and without morphological disturbance) per age bin. The declining trend in $\zeta^{*}$ means that the agreement between visual and automated classification weakens with the starburst age. The reason for this is twofold: 1) the visual classification was computed by combining individual classifications obtained by different reviewers and is therefore affected by the differences in opinions between them; 2) the automated approach is designed to measure the degree of asymmetry of the tidal features present in galaxies, while the visual approach classifies galaxies according to the presence/absence of such features only. The second statement is supported by the declining trend in $A_{S}$ as a function of the starburst age for the morphologically disturbed (post-)starburst galaxies (Figure \ref{fig:bulge_distundist}a).

\subsection{Merger, starburst, post-merger, post-starburst, ...red sequence?}\label{sec:discussion_galevo}
In this section we discuss the implications of our findings on the role of galaxy mergers in inducing starburst events in massive star-forming galaxies, and that of (post-)starburst galaxies in the build up of the red sequence. 

\subsubsection{From merger to (post-)starburst}\label{sec:discussion_galevo1}
\begin{figure}
    \subfigure{a) (Post-)starburst galaxies with disturbed morphologies (r-band)}
    {\includegraphics[width=1.0\columnwidth]{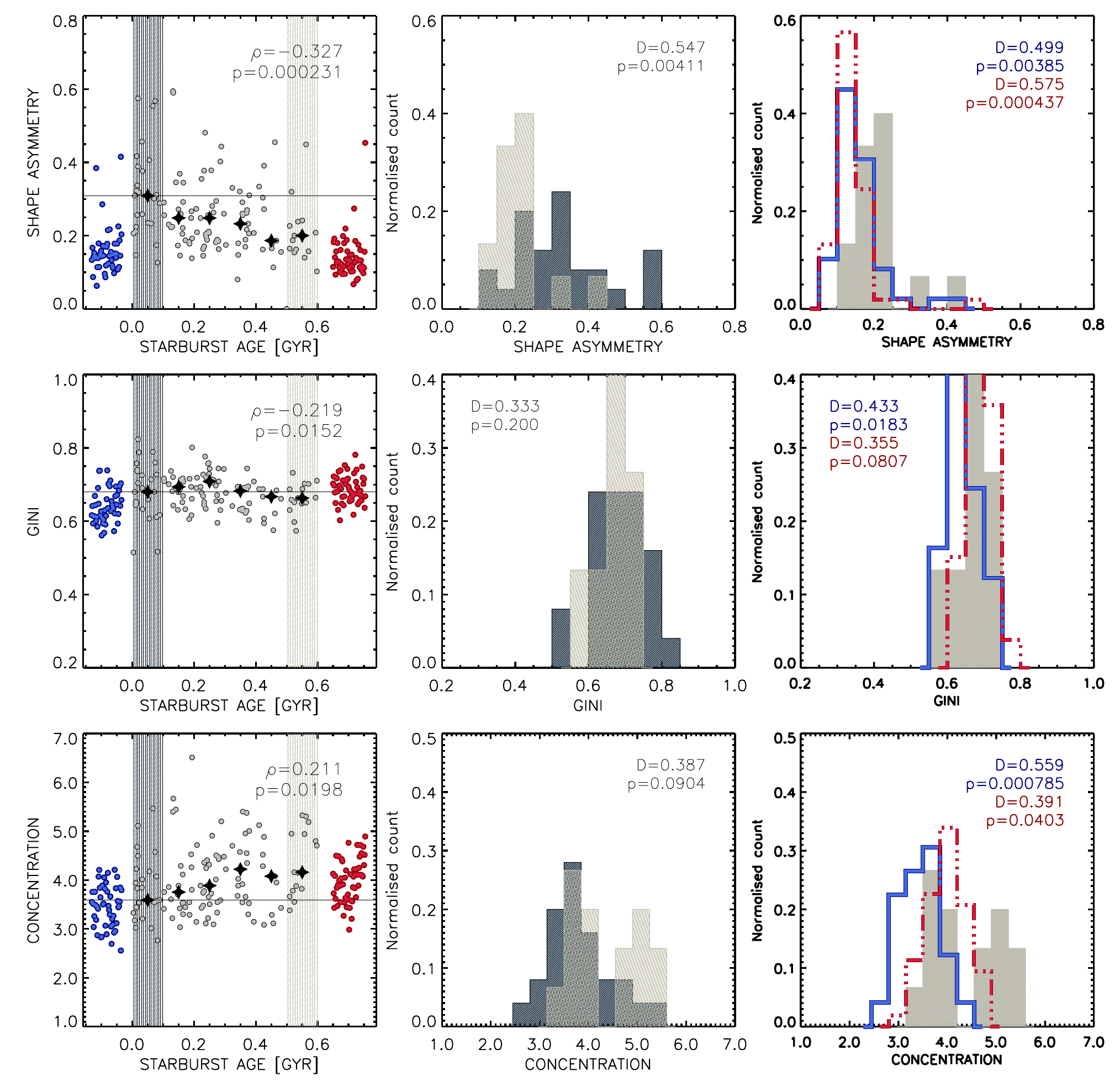}\label{fig:first_sub}} \\
    \subfigure{b) (Post-)starburst galaxies with regular morphologies (r-band)}
    {\includegraphics[width=1.0\columnwidth]{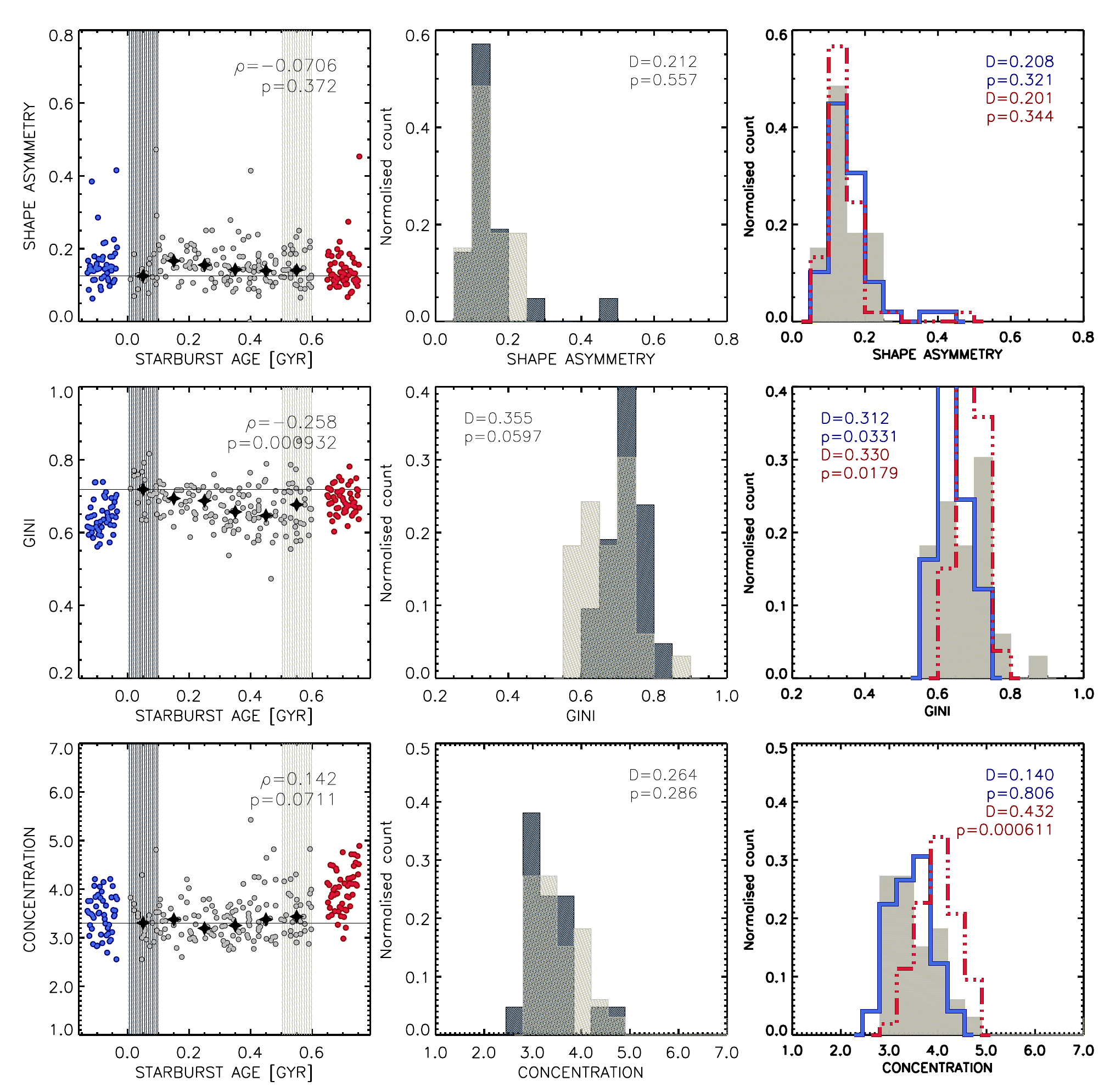}\label{fig:second_sub}} \\
     \subfigure{c) (Post-)starburst galaxies with regular morphologies (g-band)}
    {\includegraphics[width=1.0\columnwidth]{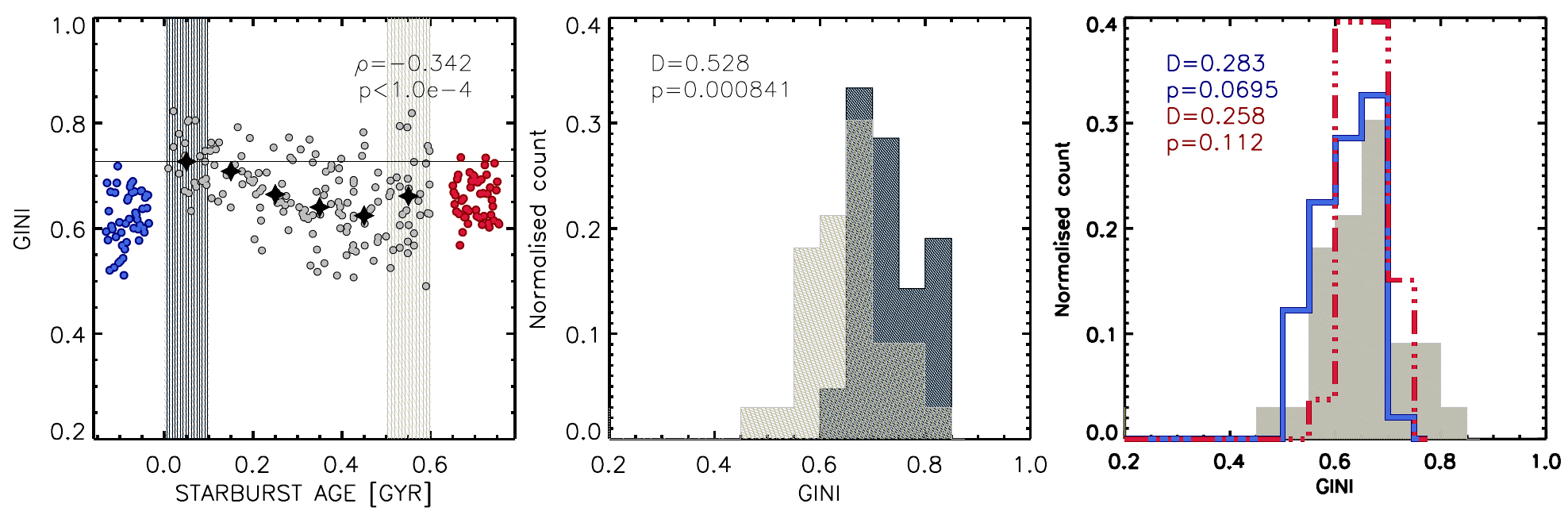}\label{fig:third_sub}} \\
    \caption{Measures of $A_{S}$, $G$ and $C$ for visually selected subsets of the (post-)starburst sample: a) morphologically disturbed galaxies, b) galaxies with regular morphologies. Additionally, in c), we show the values of $G$ measured in the $g$-band for the subset of (post-)starburst galaxies with regular morphologies. For legend, see Figures \ref{fig:nCGM20_age} and \ref{fig:SAAoAs_age}.}
    \label{fig:bulge_distundist}
\end{figure}

The results of this study are in a general agreement with previous works on the presence of tidal features in the structure of some galaxies with post-starburst stellar populations \citep{Zabludoff et al. 1996, Blake et al. 2004, Goto 2005, Yang et al. 2008}. We do not make direct quantitative comparisons due to our different selection criteria, and image limiting magnitudes, and consequently, different sample properties, compared to previous studies. 
Here, we consider an evolutionary sample of starburst and post-starburst galaxies and trace the changes in their morphology as a function of the starburst age, while the results of the previous studies refer to generally older post-starburst galaxies with no spectroscopic evidence of ongoing star-formation. 
We find that at least $45\%$ of the youngest starburst galaxies in our sample show disturbed morphological features, and that this fraction decreases with the starburst age to about $25-30\%$ at $t_{SB}>0.3$ Gyr (left panel, Figure \ref{fig:visauto}). This latter fraction is consistent with the fraction of ordinary blue-sequence galaxies that are visually classified as being morphologically disturbed (Figure \ref{fig:visual}), and therefore the morphological disturbance may not be directly related to the starburst event.
We also find a declining trend in $A_{S}$ in morphologically disturbed galaxies, with the youngest starbursts showing a range of values berween $\sim0.2$ and 0.6, and most galaxies with $t_{SB}>0.4$ Gyr having $A_{S}\sim0.2$ (Figure \ref{fig:bulge_distundist}a). This provides a qualitative fit to the galaxy merger picture emerging from numerical simulations where, after the coalescence of the two progenitors, the morphological disturbance of the system gradually fades away leading to formation of a morphologically undisturbed remnant (see e.g. \citealt{Lotz et al. 2008}).

To investigate the role of galaxy mergers in inducing starburst events in massive star-forming galaxies, we must focus on galaxies which have experienced a very recent starburst, so that at the time of observations their tidal features have not begun to vanish. Consequently, we consider the youngest age-bin in Figure \ref{fig:visauto} (left panel) to deduce that, based on the investigated sample, we can attribute at least 45\% of the (post-)starburst signatures to galaxy mergers. The remaining 55\% could be a result of 1) either galaxy mergers/interactions under conditions that fail to produce tidal signatures in the remnants' morphologies or, 2) some other internal processes. The fact that 30\% of our control sample of blue-sequence galaxies are also visually identified as being morphologically disturbed, additionally indicates that either galaxy mergers in the local Universe can cause significant morphological disturbance without inducing a central starburst, or morphological disturbance is not a good predictor of a past merger event.

In future work, we will extend our analysis to a larger sample of galaxies with young starbursts, more representative of the entire galaxy population in the nearby Universe. We will also aim to constrain the upper limit on the fraction of (post-)starburst galaxies which have originated from a merger by considering galaxy mergers found in cosmological simulations.

\subsubsection{From (post-)starburst to quiescence}\label{sec:discussion_galevo2}

The values of $G$ obtained for galaxies with the youngest starbursts ($t_{SB}<0.1$ Gyr) are as high or even higher than those found for the continuously star-forming and quiescent galaxies, and they tend to decrease with the starburst age. In the previous section, we attributed this effect primarily to the decaying nuclear starburst but we also noted that contributions from faint tidal features in galaxies may increase the corresponding values of $G$. Although our analysis of the test sample suggests that the presence of faint tidal features should have a marginal effect on $G$ in (post-)starburst galaxies, here, we address this concept again by investigating the behaviour of the Gini index in the morphologically disturbed and undisturbed subsets of our (post-)starburst sample separately.

In Figure \ref{fig:bulge_distundist}b we present the evolution of $G$ for (post-)starburst galaxies with regular morphologies as identified by visual inspection. In this case, we still observe a declining trend in $G$ with starburst age ($\rho\sim-0.26$, $p\sim10^{-4}$), however slightly less pronounced than that found for the total sample ($\rho\sim-0.29$, $p<10^{-4}$). This implies that the contribution from the tidal features is indeed not consequential and that the time-evolution of the inequality in the light distribution within (post-)starburst galaxies is predominantly an effect of the decaying starburst. This is further supported by Figure \ref{fig:bulge_distundist}c, where we consider the values of $G$ computed using images in the $g$-band.
In this regime, there is more light from younger stars with shorter lifetimes than in the case of the $r$-band, and therefore, the effects of the decaying starburst should be more apparent. As expected, we find a stronger declining trend in $G$ as a function of the starburst age in the $g$-band ($\rho\sim-0.34$, $p<10^{-4}$). The observed trends in the values of $G$ suggest that the photometric signatures of the most recent starburst, in both the $r$-band and the $g$-band, fade away after $\sim0.5$ Gyr since the starburst event.

Our analysis shows that the (post-)starburst galaxies typically have structural properties characteristic of early-type disks (reflected in the values of $n$, $C$ and $M_{20}$), with values similar to those found for the star-forming blue-cloud galaxies and somewhat lower than those measured for the quiescent red-sequence galaxies in our control samples. The average structural properties of the sample are largely a result of our sample selection, where we only considered galaxies with high stellar surface mass densities. More interestingly, the lack of evolution in these structural parameters with starburst age, and in particular the values of $n$, $C$ and $M_{20}$ for the oldest (post-)starbursts, leads us to conclude that the (post-)starburst galaxies do not attain the highly concentrated structure typical for red-sequence galaxies within the first 0.6 Gyr following a merger. This does not necessarily imply that they are not heading toward the red sequence; perhaps further structural evolution is required for the final transition to occur. A similar conclusion was found in a recent study of a subset of the (post-)starburst galaxies used in this work, by \citet{Rowlands et al. 2015}, who found a substantial amount of gas remaining in these galaxies, that could sustain star formation for at least 0.5-1 Gyr after the starburst, suggesting that the post-starburst galaxies are not yet fully quenched.
 
We note the following caveats in our conclusions. The galaxies in our control sample of red-sequence galaxies to which we compare, have not necessarily quenched their star formation recently and could have undergone some structural evolution while residing on the red sequence. Furthermore, present-day red-sequence galaxies that experienced their star-formation quenching earlier in the history of the Universe, when the physical conditions were significantly different to the present day, may not be representative of the `future' red-sequence which will form once the residual star-formation within the present-day (post-)starburst galaxies have ceased. Further study of (post-)starburst galaxies at higher redshifts, including those with $t_{SB}>0.6$ Gyr, as well as a more extended sample of red-sequence galaxies, and a quantitative comparison with previous studies are underway.

\section{Summary}
It is thought that the observed bimodality in galaxy properties could be a result of a transformation from gas-rich actively star-forming galaxies to passively evolving galaxies devoid of gas, following some star-formation quenching event (e.g. Bell et al. 2004, Faber et al. 2007). In this work, we focused on one of the possible quenching routes, involving a transition through a post-starburst phase subsequent to a gas-rich merger event (e.g. Sanders et al. 1988, Wild et al. 2009, Yesuf et al. 2014).
Our study involved the morphological and structural analysis of an evolutionary sample of 335 bulge-galaxies with the strongest (post-)starburst signatures within the SDSS DR7, with equal number of galaxies per unit starburst age ($t_{SB}$) with $t_{SB}<0.6$ Gyr. The unique sample properties provided an opportunity for a statistical study of a sequence of galaxies evolving through the short-lasting phase following a nuclear starburst. For the analysis, we used standard measures of galaxy structure found in the literature as well as a newly introduced morphological indicator designed to trace faint asymmetric tidal features in the outskirts of the galaxies. Below, we list the main results of our study and their implications.
 \begin{itemize} 
 \item The new measure of morphological asymmetry introduced in this work provides a robust method for probing asymmetric tidal features in the outskirts of galaxies and allows for an effective detection of galaxies in late stages of a merger. Within our test sample, the shape asymmetry criterion $A_{S}>0.2$ identifies 95$\%$ (19/20) of galaxies with faint tidal features and $60\%$ (6/10) of those with low/moderate morphological disturbance. Applied to our (post-)starburst sample, $A_{S}>0.2$ detects $45\%$ of the galaxies with the youngest starbursts ($t_{SB}<0.1$ Gyr), a fraction which is in a remarkable agreement with that obtained by visual inspection of the images.
 \item The fraction of (post-)starburst galaxies which show features characteristic of post-mergers declines with the starburst age. For the morphologically disturbed (post-)starburst galaxies, the measured values of $A_{S}$ decrease with the starburst age, with $A_{S}\sim0.2$ for most galaxies with starbursts older than 0.5 Gyr.These trends fit qualitatively to the galaxy merger picture, where the morphological disturbance of the merger remnant gradually fades away as the remnant evolves through the dynamically-cold post-merger phase (see e.g. \citealt{Lotz et al. 2008}). 
\item Assuming that morphological disturbance in galaxies is an indicator of a past merger, both visual classification and the automated approach (using the shape asymmetry) suggest that at least $45\%$ of the youngest (post-)starburst galaxies in the sample ($t_{SB}<0.1$ Gyr) have originated in a merger.
\item The insignificant age-evolution of their central concentration, light profiles or the extent of the brightest regions, suggests that (post-)starburst galaxies do not attain the structure of the fully-quenched galaxies within the first 0.6 Gyr after the starburst. Rather, their structure bears more resemblance to that of the continuously star-forming galaxies. Further structural evolution is required before the post-starburst galaxies can become comparable to galaxies populating the present-day red sequence.
 \end{itemize}

\section*{Acknowledgments}
The authors would like to thank the anonymous referee for thorough reading of the manuscript and useful comments and suggestions, which led to clarification of several important points in the text.

MMP and VW acknowledge support from the  European Career Reintegration Grant Phiz-Ev (P.I. V. Wild). VW, KR and JM-A acknowledge support from the European Research Council Starting Grant SEDMorph (P.I. V. Wild). CJW acknowledges support through the Marie Curie Career Integration Grant 303912. PHJ acknowledges the support of the Academy of Finland grant 1274931.

Funding for the SDSS and SDSS-II has been provided by the Alfred P. Sloan Foundation, the Participating Institutions, the National Science Foundation, the U.S. Department of Energy, the National Aeronautics and Space Administration, the Japanese Monbukagakusho, the Max Planck Society, and the Higher Education Funding Council for England. The SDSS Web Site is http://www.sdss.org/.

The SDSS is managed by the Astrophysical Research Consortium for the Participating Institutions. The Participating Institutions are the American Museum of Natural History, Astrophysical Institute Potsdam, University of Basel, University of Cambridge, Case Western Reserve University, University of Chicago, Drexel University, Fermilab, the Institute for Advanced Study, the Japan Participation Group, Johns Hopkins University, the Joint Institute for Nuclear Astrophysics, the Kavli Institute for Particle Astrophysics and Cosmology, the Korean Scientist Group, the Chinese Academy of Sciences (LAMOST), Los Alamos National Laboratory, the Max-Planck-Institute for Astronomy (MPIA), the Max-Planck-Institute for Astrophysics (MPA), New Mexico State University, Ohio State University, University of Pittsburgh, University of Portsmouth, Princeton University, the United States Naval Observatory, and the University of Washington.

\appendix

\section{The test sample}\label{app:test}
\begin{figure*}
  \centering  
  \includegraphics[scale=0.8]{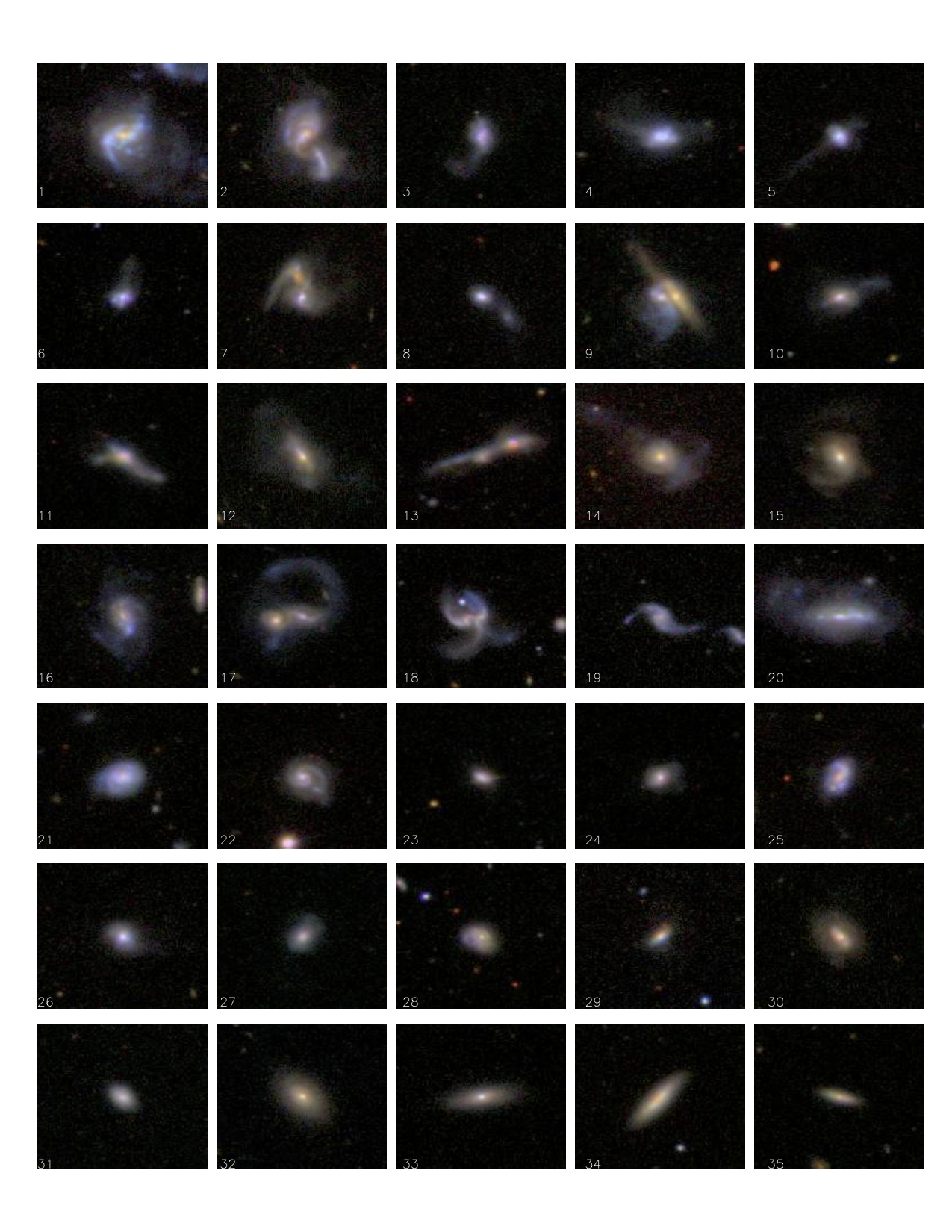}
  \caption{Three-colour images of the post-starburst galaxies in the test sample, including objects with: high morphological disturbance ($1-20$), low morphological disturbance ($21-30$) and regular morphology ($31-35$).}
  \label{fig:morphsample1}
\end{figure*}
\begin{figure*}
  \centering  
  \includegraphics[scale=0.8]{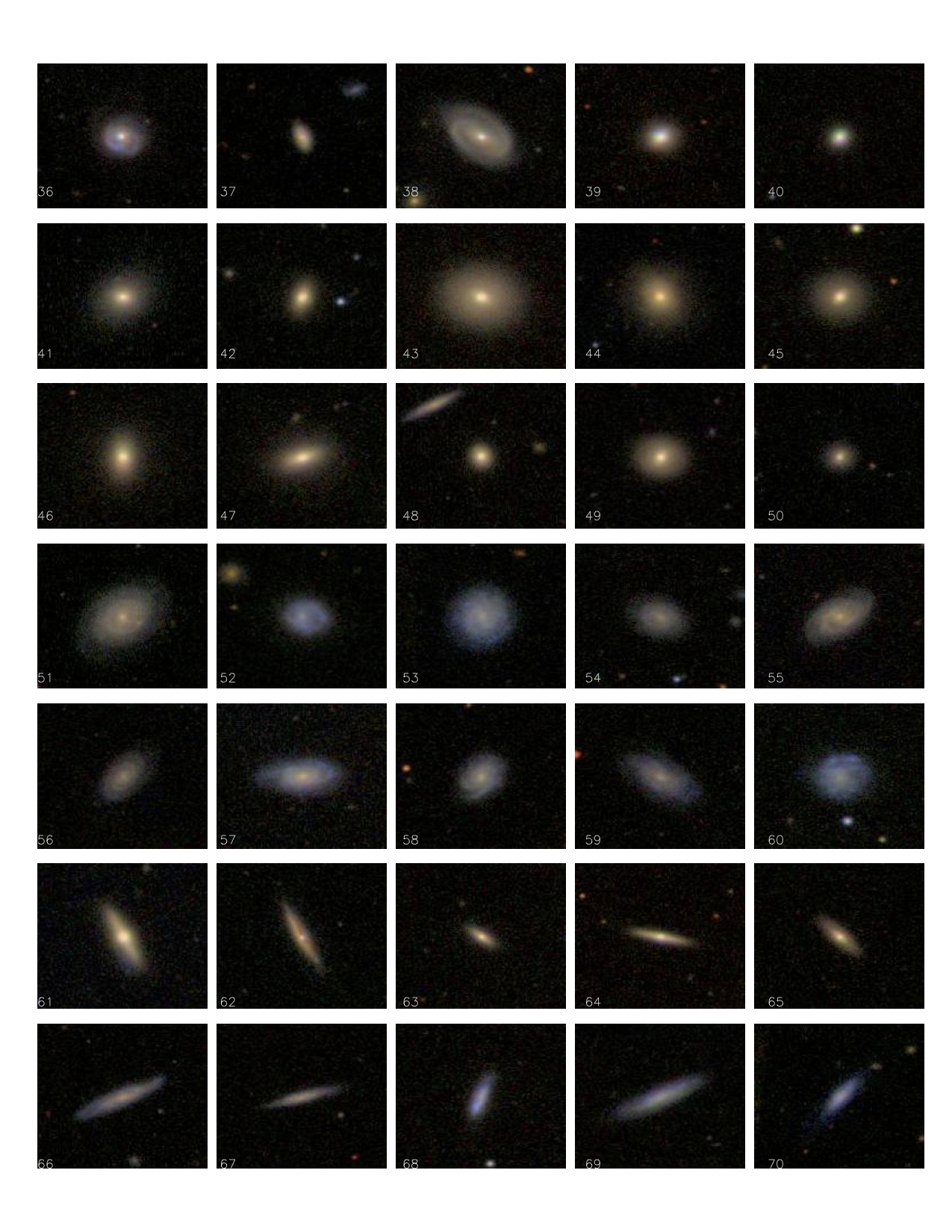}
  \caption{Three-colour images of the remaining objects in the test sample: post-starburst galaxies with regular morphology (36-40) and normal galaxies, including early-types ($41-50$; $61-65$) and late-types ($51-60$; $67-70$).}
  
  \label{fig:morphsample2}
\end{figure*}

\label{lastpage}

\end{document}